\documentclass[prb,twocolumn,showpacs,preprintnumbers,amsmath,amssymb,superscriptaddress]{revtex4-1}

\usepackage{graphicx}
\usepackage{dcolumn}
\usepackage{color}

\def\a{\alpha}
\def\b{\beta}

\def\iomn{i\omega_n}
\def\inun{i\nu_n}
\def\vq{{\bf q}}
\def\vk{{\bf k}}

\def\t{\mbox{tr}}

\def\cG0{{\cal G}_0}
\def\GRRp{G}
\def\WRRp{W}
\def\GRR{G^{loc}}
\def\WRR{W^{loc}}

\def\cU{{\cal U}}
\def\cG{{\cal G}}
\def\GH{{G_H}}
\def \t2g{t$_{2g}$}
\allowdisplaybreaks
\newcommand{\beq}{\begin{equation}}
\newcommand{\eeq}{\end{equation}}
\newcommand{\beqa}{\begin{eqnarray}}
\newcommand{\eeqa}{\end{eqnarray}}

\newcommand{\fref}[1]{Fig.~\ref{#1}}

\newcommand{\jmbra}[1]{\ensuremath{\langle #1|}}
\newcommand{\jmket}[1]{\ensuremath{|#1\rangle}}
\newcommand{\GW}{{\it GW}}

\newcommand{\svek}[1]{%
        {\mathbf #1}}

\newcommand{\out}[1]{%
        {}}

\renewcommand{\Im}{\mathrm{Im}}
\renewcommand{\Re}{\mathrm{Re}}

\begin{document}
\title{
Asymmetric band widening by screened exchange
competing with local correlations in 
SrVO$_3$: new surprises on an old compound from
combined {\it GW} and dynamical
mean field theory {\it GW}+DMFT
}

\author{Jan~M.~Tomczak}
\affiliation{Institute of Solid State Physics, Vienna University of Technology, A-1040 Vienna, Austria}
\author{M.~Casula}
\affiliation{CNRS and Institut de Min{\'e}ralogie et de Physique des Milieux condens{\'e}s, Universit{\'e} Pierre et Marie Curie, case 115, 4 place Jussieu, 75252, Paris cedex 05, France}
\author{T.~Miyake}
\affiliation{Nanosystem Research Institute, AIST, Tsukuba 305-8568, Japan}
\author{S.~Biermann}
\affiliation{Centre de Physique Th{\'e}orique, Ecole Polytechnique, CNRS-UMR7644, 91128 Palaiseau, France}


\begin{abstract}
The very first dynamical 
implementation of the combined {\it GW} and dynamical mean field scheme
``{\it GW}+DMFT'' for a real
material was achieved recently [J.M. Tomczak et al., Europhys.
Lett. 2012], 
and applied to the ternary transition metal oxide SrVO$_3$.
Here, we review and extend that work, giving not only a detailed
account of full {\it GW}+DMFT calculations, but also
discussing and testing simplified approximate schemes.
We give insights into the nature of exchange and correlation 
effects:
Dynamical renormalizations in the Fermi liquid regime of SrVO$_3$ 
are essentially local, and nonlocal correlations mainly act to
screen the Fock exchange term.
The latter substantially widens the quasi-particle band structure,
while the band narrowing induced by the former is accompanied
by a spectral weight transfer to higher energies.
Most interestingly, the exchange broadening is much more
pronounced in the unoccupied part of spectrum.
As a result, the {\it GW}+DMFT electronic structure of SrVO$_3$
resembles the conventional 
density functional based dynamical mean field (DFT+DMFT)
description for occupied states, 
but is profoundly modified in the empty
part. 
Our work leads to a reinterpretation of inverse
photoemission spectroscopy (IPES) data. Indeed, we assign a prominent peak 
at about 2.7 eV dominantly to e$_g$ states, rather than to
an upper Hubbard band of t$_{2g}$
character.
Similar surprises can be expected for other
transition metal oxides, calling for more detailed investigations
of the conduction band states. 
\end{abstract}
\date{\today}
\maketitle

\section{Introduction}

Within the last decade,
a new research field has developed at the interface of 
many-body theory and first principles electronic structure 
calculations. The aim is the construction of materials-specific
parameter-free many-body theories that preserve the {\it ab initio}
nature of density functional based methods,  
but incorporate at the same time a 
many-body description of Coulomb interactions beyond the
independent-electron picture into 
computational approaches for
spectroscopic or finite-temperature properties.

Historically, the first non-perturbative electronic structure 
techniques for correlated materials evolved from many-body 
treatments of the multi-orbital Hubbard Hamiltonian with
realistic parameters.
The general strategy of these so-called ``LDA++'' approaches
\cite{PhysRevB.57.6884, 0953-8984-9-35-010}
(for reviews see, e.g.,
\cite{biermann_ldadmft,held_psik,anisimovbook,vollkot})
consists in the extraction of the 
parameters of a many-body Hamiltonian from first principles 
calculations and then solving the problem by many-body 
techniques. 
In practice, this procedure has met tremendous success
in the description of the electronic structure of correlated
materials, for a wide range of materials, from
transition metals \cite{licht_katsnelson_kotliar,biermann_akw}, 
their oxides \cite{PhysRevLett.90.096401,pavarini:176403,PhysRevLett.86.5345,keller:205116,poter_v2o3,PhysRevB.72.155106,tomczak_v2o3_proc,nekrasov:155112,PhysRevB.71.153108,PhysRevLett.104.047401,
PhysRevB.85.115136,PhysRevB.86.195136,PhysRevB.86.184413,PhysRevB.85.035124,PhysRevLett.104.226401,PhysRevLett.104.086402,PhysRevLett.99.126402,PhysRevLett.102.146402,PhysRevLett.110.267204,biermann:026404,me_psik,optic_epl}, 
sulphides \cite{PhysRevLett.94.166402, lechermann:085101}, 
or silicides \cite{jmt_fesi,jmt_hvar}, 
to $f$-electron 
compounds \cite{JPSJS.77SC.99,jmt_cesf, deltaPu, PhysRevLett.87.276404}.
More recently, iron pnictide compounds 
(see e.g.\ 
Refs.~\onlinecite{PhysRevLett.100.226402,
PhysRevB.80.092501,0953-8984-21-7-075602,
PhysRevB.80.085101,PhysRevB.82.064504,
PhysRevLett.104.197002,PhysRevB.81.220506,PhysRevB.85.094505,werner_bfa})
or spin-orbit materials \cite{PhysRevLett.107.266404} have come into the
focus of many-body electronic structure calculations, emphasizing 
the need for fully {\it ab initio} techniques, including a 
first principles description of the effective Coulomb interactions.
The challenge here is an accurate description of screening
of low-energy interactions by high-energy degrees of 
freedom, as well as the screening of local interactions by
nonlocal charge fluctuations \cite{ayral_gwdmft,PhysRevB.87.125149,
PhysRevLett.110.166401}.

Despite the tremendous success of LDA++ schemes, 
one should be aware of the fact that the ambiguities in the
construction of the Hamiltonian are not limited to the many-body
part: not even the use of the Kohn-Sham band structure of DFT 
as a starting Hamiltonian has a direct microscopic justification 
beyond heuristic arguments.
Though renormalization group techniques suggest that in 
many cases the relevant low-energy effective Hamiltonian
can indeed be cast into a generalized (multi-orbital)
Hubbard form, in practice neither the precise form nor 
the parameters can be derived directly from the Coulomb
Hamiltonian in the continuum.
In this sense, the construction of an ``LDA++'' Hamiltonian
amounts to a rather {\it ad hoc} combination of a Kohn-Sham
Hamiltonian and multi-orbital Hubbard (and Hund) interaction terms for
a subset of ``correlated orbitals''.
Conceptually, there is moreover a mismatch arising from the fact 
that the full long-range Coulomb interactions enter the
one-particle part of the Hamiltonian
(even if only in a mean-field fashion), while in the many-body
part they are replaced by effective local interactions acting
only in a low-energy subspace.
This has two consequences. The first -- well-known one -- is
related to the double counting correction: 
Correlation effects accounted for in the exchange-correlation potential of DFT
have to be subtracted. Yet, a 
microscopically motivated definition 
of this term is, even on a conceptual level, impossible.
The second one is more subtle, and has only recently started
to receive some attention: in fact, the same processes that
screen the effective Coulomb interactions
are also responsible for renormalizations of the
one-body part of the Hamiltonian. This can be understood
from an analysis of screening as resulting from coupling of the
electrons to bosonic excitations, such as plasmons, particle-hole 
excitations or more complex many-body processes.
The diagonalization of the corresponding electron-boson
Hamiltonian results in fermionic quasi-particles (``electronic
polarons'') corresponding to electrons dressed by their 
screening bosons, and thus having heavier masses.
This mass enhancement corresponds to an effective renormalization
of their kinetic energy, and hence of the one-body part of the
Hamiltonian. This kind of effect has recently been demonstrated
explicitly \cite{casula_effmodel} on the basis of the constrained random phase
approximation (cRPA), which allows for an explicit (yet approximate)
estimation of dynamical Hubbard interactions in
solids \cite{PhysRevB.70.195104}. 
The corresponding one-body renormalizations
have been investigated in the framework of dynamical mean field
theory (DMFT) for SrVO$_3$ \cite{PhysRevB.85.035115} and 
BaFe$_2$As$_2$ \cite{werner_bfa},
and a low-energy effective Hamiltonian comprising these
renormalizations has been derived in Ref.~\onlinecite{casula_effmodel}.

In addition to these effects related to the long-range
nature of the Coulomb interactions and the resulting
quantum dynamical screening, in practice, yet another 
difficulty arises when proceeding to (approximate) many-body
solutions of the multi-orbital Hubbard Hamiltonian.
Indeed, while the construction of the one-body part of the
Hamiltonian (within DFT) naturally puts the electronic density 
at the center of the attention, many-body theory is most
readily formulated within a Green's function language.
This mismatch in language is the final capstone that ensures
that matching contributions between the effective one-body 
Hamiltonian and the many-body terms are truly impossible to identify.

Ideally, the desired specifications of new many-body
electronic structure techniques beyond ``LDA++'' approaches
can thus be summarized in three main requirements:
\begin{itemize}
\item The theory should be entirely formulated in the Green's function
language, even at the one-body level.
\item The theory should deal directly with the long-range
Coulomb interactions, and any effective local ``Hubbard-like''
interactions should arise only as intermediate auxiliary
quantities.
\item At the same time, the theory should retain the non-perturbative
character of dynamical mean field theory, thus avoiding limitations
due to a truncation of the perturbation series. This latter point
is essential to ensure the scheme to be equally appropriate in
the weak, strong and intermediate coupling regimes.
\end{itemize}
The combination of Hedin's {\it GW} approximation -- many-body
perturbation theory to first order in the screened Coulomb
interaction $W$ -- and dynamical mean field theory meets
these criteria. Such a scheme was proposed a decade ago
\cite{PhysRevLett.90.086402},
based on the construction of the free energy of a solid as
a functional of the Green's function $G$ and $W$.

Only very recently have practical implementations 
for real materials 
been achieved\cite{jmt_svo,PhysRevLett.110.166401}
that go beyond simple static approximation
schemes \cite{PhysRevLett.90.086402,0953-8984-17-48-010,PhysRevB.88.165119}.
The reason was the necessity
of dealing with frequency-dependent interactions at the DMFT
level, which has remained a major bottleneck until recently.
Recent advances in Monte Carlo techniques\cite{PhysRevLett.104.146401} and the invention of a reliable
cumulant-type scheme, the ``Bose factor ansatz'' \cite{PhysRevB.85.035115}, 
have unblocked the situation:
two calculations within {\it GW}+DMFT taking into account 
dynamical interactions have been achieved recently,
for SrVO$_3$ \cite{jmt_svo} and for systems of adatoms on surfaces
\cite{PhysRevLett.110.166401}.
In this work, we review and extend the former calculations, 
giving a detailed account of fully dynamical {\it GW}+DMFT calculations
for SrVO$_3$. The paper is organised as follows:
In Sect.~II, we give an extensive summary of the concepts of the
combined {\it GW}+DMFT scheme and discuss aspects of its practical 
implementation, in particular related to the Bose factor ansatz.
Furthermore, we devote an extensive discussion to the question of
how to treat multi-orbital materials: we propose that for ligand 
and conduction band shells a perturbative treatment might be
sufficient, and show how such a procedure can be combined with
the non-perturbative DMFT treatment of the low-energy correlated
shells.
In Sect.~III, we review the electronic structure of our target
compound, pointing out problems left open within conventional
LDA++ schemes. Section IV presents the results of fully
dynamical {\it GW}+DMFT calculations, in comparison to {\it GW} calculations,
LDA+DMFT with static and dynamic interactions, and to
simplified combinations of {\it GW} and DMFT which allow for a detailed
analysis of the importance of the different terms entering
the theory. We discuss the implications of our results in 
Section V, before arriving at our conclusions in Sect.~VI.

\section{The ``{\it GW}+DMFT'' Methodology}

\subsection{Overview}

The starting point of the {\it GW}+DMFT scheme
is Hedin's {\it GW} approximation (GWA)\cite{hedin}, 
in which the
self-energy of a quantum many-body system is obtained from a frequency convolution 
(or product in time) of
the Green's function G with the screened Coulomb interaction 
$W=\epsilon^{-1}V$. 
The dielectric function $\epsilon $, which screens the bare Coulomb
potential $V$, is -- within a pure {\it GW} scheme -- obtained from the random
phase approximation. The {\it GW}+DMFT scheme, as proposed in \cite{PhysRevLett.90.086402},  
combines the first principles description of screening inherent in {\it GW}
methods with the non-perturbative nature of DMFT, where local quantities
such as the local Green's function are calculated to all orders 
in the interaction from an
effective reference system 
(\textquotedblleft impurity model\textquotedblright )%
\footnote{%
The notion of locality refers to the use of a specific basis set of
atom-centered orbitals, such as muffin-tin orbitals, or atom-centered Wannier
functions.}. 
In DMFT, one imposes a self-consistency condition for the one-particle
Green's function, namely, that its on--site projection equals the impurity 
Green's function. In {\it GW}+DMFT, the self-consistency requirement is
generalized to encompass also two-particle quantities, namely, the local
projection of the screened interaction is required to equal the  
impurity screened interaction. This 
in principle promotes the Hubbard U from an adjustable parameter in
DMFT techniques to a self-consistent auxiliary function that incorporates
long-range screening effects in an {\it ab initio} fashion. 
Indeed, as already alluded to above, 
not only higher energy degrees
of freedom can be downfolded into an effective dynamical interaction,
but one can also aim at incorporating nonlocal screening effects
into an effective dynamical $\mathcal{U}(\omega)$.
The theory is then free of any Hubbard {\it parameter}, and the
interactions are directly determined from the full long-range
Coulomb interactions in the continuum.

From a formal point of view, the {\it GW}+DMFT method, as introduced in \cite{PhysRevLett.90.086402}%
\footnote{see also the related scheme of Ref.~\onlinecite{PhysRevB.66.085120}
and the comparison to {\it GW}+DMFT in Ref.~\onlinecite{PhysRevB.87.125149}}, 
corresponds to a specific 
approximation to the correlation 
part of the free energy of a solid, expressed as a functional of the Green's
function G and the screened Coulomb interaction W: the nonlocal part is
taken to be the first order term in $W$, while the local part is calculated
from a local impurity model as in (extended) dynamical mean field theory.
This leads to a set of self-consistent equations for the Green's function 
$G$, 
the screened Coulomb interaction $W$, the self-energy $\Sigma $ and the
polarization $P$ \cite{gwdmft_proc2,gwdmft_proc1}.
Specifically, the self-energy is obtained as $\Sigma
=\Sigma^{local}+\Sigma^{nonlocal}_{GW}$, where the local part 
$\Sigma_{local}$ is derived from the impurity model. 
In practice, however, the calculation of a self-energy for
(rather delocalized) s- or p-orbitals has never been performed 
within DMFT, and it appears to be more physical to approximate 
this part also by a {\it GW}-like expression. 
For these reasons Ref.~\onlinecite{jmt_svo} proposed a practical scheme,
in which only the local part of the self-energy of the \textquotedblleft
correlated\textquotedblright\ orbitals is calculated from the impurity model
and all other local and nonlocal components are
approximated by their first order expressions in $W$. 

In the following subsections, we first briefly summarize
the functional formulation of the {\it GW}, DMFT and {\it GW}+DMFT schemes
from a general point of view (section \ref{unified}). The
corresponding {\it GW}+DMFT equations are summarized in appendix A. 
Sections \ref{orbital-separation} and \ref{orbsepeqns} are
devoted to the ``orbital-separated scheme''
implemented for SrVO$_3$, defining the equations solved
in practice.
We then review the dynamic atomic limit approximation for
the solution of dynamical impurity models (section \ref{dala}),
while section \ref{technical} summarizes
some technicalities.

\subsection{Unified view on {\it GW}, DMFT, and {\it GW}+DMFT}
\label{unified}

Within the Born-Oppenheimer approximation, the electronic
many-body states in a solid are determined by the eigenstates of the
Coulomb Hamiltonian
\begin{eqnarray}
H = H_{kin} + H_{pot} + H_{ee}
\end{eqnarray}
where the first two terms denote the kinetic energy part
and one-body potential created by the ions respectively.
The last term,
$H_{ee} = \sum_{n m n^\prime m^\prime} v_{n m n^\prime m^\prime} 
a^{\dagger}_{n} a^{\dagger}_{m} a_{m^\prime} a_{n^\prime}$,
with 
$v_{n m n^\prime m^\prime} = \langle
n m | \frac{1}{|r-r^{\prime}} | n^\prime m^\prime \rangle$
the matrix elements of the Coulomb interaction in the 
continuum, denotes the electron-electron interaction.

Following Almbladh {\it et al.}\cite{Almbladh_1999},
the free energy of a solid can be formulated as a functional $\Gamma[G,W]$
of the Green's function $G$ and the screened Coulomb interaction
$W$ of the solid. The latter is defined as the correlation function 
of bosonic excitations corresponding to density fluctuations,
that is, in mathematical terms, as the propagator of the 
Hubbard-Stratonovich field decoupling the Coulomb interaction term.
The {\it GW} method, dynamical mean field theory
and the combined {\it GW}+DMFT scheme
can then be viewed as different
approximations to this $\Gamma[G,W]$ functional. 

The functional $\Gamma$ can trivially be split into a Hartree
part $\Gamma_H$ and a many body correction $\Psi$, which
contains all corrections beyond the Hartree approximation~:
$\Gamma =  \Gamma_H + \Psi$. The Hartree part can be given
in the form
\begin{eqnarray}
\Gamma_H[G,W]
&=&Tr\ln G-Tr[({G_H}^{-1}-G^{-1})G]
\nonumber
\\
&-&\frac{1}{2}Tr\ln W+\frac{1}%
{2}Tr[({V_q}^{-1}-W^{-1})W]
\label{LW}%
\end{eqnarray}
with $G_H$ being the Hartree Green's function, and $V_q$ the
Fourier transform of the bare Coulomb interaction.
The $\Psi$-functional
is the sum of all skeleton
diagrams that are irreducible with respect to both one-electron
propagator and interaction lines.
 $\Psi[G,W]$ has the following properties:
\begin{eqnarray}\label{eq:def_sigma_P} 
\frac{\delta \Psi}{\delta G} = \Sigma^{xc}
\nonumber
\\
\frac{\delta \Psi}{\delta W} = P.
\end{eqnarray}
The $\Psi$ functional was first derived in \cite{Almbladh_1999}.
A detailed discussion in the context of extended DMFT can
be found in Ref.~\onlinecite{PhysRevB.63.115110}, while Refs.~\onlinecite{PhysRevLett.90.086402, gwdmft_proc2,gwdmft_proc1}
view it from the {\it GW}+DMFT point of view. 

An elegant derivation 
(see e.g.~\onlinecite{gwdmft_proc2,gwdmft_proc1, PhysRevLett.110.166401})
of the Almbladh free energy functional
is obtained through a Hubbard Stratonovich
decoupling of the interaction term by a bosonic field $\phi$,
the introduction of Lagrange multipliers $\Sigma$ and $P$
imposing $\langle c c^{\dagger} \rangle$ and $\langle \phi \phi \rangle$
to equal externally chosen fermionic and bosonic
propagators $G$ and $W$, and finally a Legendre transformation
to obtain a functional of the latter two quantities.

The {\it GW} approximation consists in retaining the first order
term in the screened interaction $W$ 
only, thus 
approximating the $\Psi$-functional
by
\begin{eqnarray}
\Psi[G,W] = -\frac{1}{2} Tr(GWG).
\end{eqnarray}
We then trivially find 
\begin{eqnarray}
\Sigma^{xc} = \frac{\delta \Psi}{\delta G} = -G W
\end{eqnarray}
\begin{eqnarray}
P = \frac{\delta \Psi}{\delta W} = G G.
\end{eqnarray}
 
Extended DMFT%
\cite{Kajueter,PhysRevLett.77.3391,PhysRevB.52.10295}, 
on the other hand, would calculate all
local quantities that should be derived from this functional 
from a local
impurity model. One can thus formally write
\begin{eqnarray}\label{Psi-edmft}
\Psi\,=\,\Psi_{imp}[\GRR,\WRR].
\end{eqnarray}

The combined {\it GW}+DMFT scheme \cite{PhysRevLett.90.086402} consists in approximating
the $\Psi$ functional as a direct combination of local
and nonlocal parts from {\it GW} and extended DMFT, respectively:
\begin{eqnarray}\label{Psi}
\Psi\,=\,\Psi_{GW}^{\rm{non-loc}}[\GRRp,\WRRp]
+\Psi_{imp}[\GRR,\WRR]
\end{eqnarray}
More explicitly, the nonlocal part of the {\it GW}+DMFT
$\Psi$-functional is given by
\begin{eqnarray}
\Psi_{GW}^{\rm{non-loc}}[\GRRp,\WRRp]
= \Psi_{GW}[\GRRp,\WRRp] - \Psi_{GW}^{\rm{loc}}[\GRRp,\WRRp]
\end{eqnarray}
while the local part is taken to be an impurity model 
$\Psi$ functional.
Following (extended) DMFT, this on-site part of the functional is generated
from a local {\it quantum impurity problem}. 
The expression for its free energy functional
$\Gamma_{imp}[G_{imp},W_{imp}]$
is analogous to (\ref{LW}) with 
the Weiss field ${\cal G}$ replacing $\GH$ and the Hubbard ${\cal U}$ replacing $V$~:
\begin{eqnarray}
\Gamma_{imp}[G_{imp},W_{imp}]
&=&Tr\ln G_{imp}-Tr[({\cal G}^{-1}-G_{imp}^{-1})G_{imp}]
\nonumber
\\
&-&\frac{1}{2}Tr\ln W_{imp}+\frac{1}%
{2}Tr[({\cal U}^{-1}-W_{imp}^{-1})W_{imp}]
\nonumber
\\
&+&\Psi_{imp}\lbrack G_{imp},W_{imp} \rbrack \label{LWimp}%
\end{eqnarray}
The impurity quantities $G_{imp},W_{imp}$ can thus be
calculated from the
effective action:
\begin{eqnarray}\label{eq:action}\label{Simp}
&S&=\int d\tau 
d\tau' \left[ -
\sum 
c^{\dagger}_{L}(\tau) {\cal G}^{-1}_{LL'}(\tau-\tau')c_{L'}(\tau')
\right.
\\ \nonumber
&+&\frac{1}{2}
\left.
\sum 
:c^\dagger_{L_1}(\tau)c_{L_2}(\tau):\cU_{L_1L_2L_3L_4}(\tau-\tau')
:c^\dagger_{L_3}(\tau')c_{L_4}(\tau'): \right]
\hskip-1cm
\end{eqnarray}
where the sums run over all orbital indices $L$.
In this expression, $c_L^{\dagger}$ is a creation operator associated
with a localized orbital $L$,
and the double dots denote normal ordering (taking care of
Hartree terms).
For simplicity, we restrict the discussion to the paramagnetic case
and omit any spin indices.

The construction (\ref{Psi})
of the $\Psi$-functional is the only {\it ad hoc}
assumption in the {\it GW}+DMFT approach. The explicit form of the
{\it GW}+DMFT equations follows then directly from the functional relations
between the free energy, the Green's function, the screened
Coulomb interaction etc.
 Taking derivatives of the functional
(\ref{Psi}) as in (\ref{eq:def_sigma_P}) yields
the complete self-energy and polarization operators:
\begin{eqnarray}
\label{Sig_a}
\Sigma^{xc}(\vk,\iomn)_{LL'} &=& \Sigma_{GW}^{xc}(\vk,\iomn)_{LL'}
\\
&-& \sum_\vk \Sigma_{GW}^{xc}(\vk,\iomn)_{LL'}
+ [\Sigma^{xc}_{imp}(\iomn)]_{LL'}
\nonumber
\\
P(\vq,\inun)_{\a\b} &=& P_{GW}(\vq,\inun)_{\a\b}
\label{P_a}
\\
&-& \sum_\vq P_{GW}(\vq,\inun)_{\a\b}+ P_{imp}(\inun)_{\a\b}
\nonumber
\end{eqnarray}

Here, Greek letters indicate a two-particle basis, constructed from the localized (Wannier) basis indexed by $L$.
The {\it ad hoc} combination of the functional $\Psi$
constructed as a sum of local and nonlocal parts thus leads
to a physically attractive result:
The off-site part of the self-energy (\ref{Sig_a}) is taken from
the {\it GW} approximation, whereas the on-site part is calculated
to all orders from the dynamical impurity model.
This treatment thus goes beyond usual extended DMFT, where
the lattice self-energy and polarization are just taken
to be their impurity counterparts.
The second term in
(\ref{Sig_a}) subtracts the on-site component of the {\it GW}
self-energy thus avoiding double counting.
At self-consistency this term can be rewritten as:
\begin{equation}\label{Sig_correction}
\sum_\vk \Sigma_{GW}^{xc}(\vk,\tau)_{LL'}= - \sum_{L_1L_1'}
{W_{imp}}(\tau)_{LL_1L'L'_1} {G_{imp}}(\tau)_{L'_1L_1}
\end{equation}
so that
it precisely subtracts the contribution of the {\it GW} diagram to
the impurity self-energy. Similar considerations apply
to the polarization operator.

The general set of {\it GW}+DMFT equations to be solved self-consistently
is summarized in Appendix A. In the following, we discuss
a variant, which allows for a physically motivated cheaper
treatment of ligand and itinerant empty states.

\subsection{The ``orbital-separated'' {\it GW}+DMFT scheme}
\label{orbital-separation}

In the original {\it GW}+DMFT scheme as described in Ref.~\onlinecite{PhysRevLett.90.086402},
the $\Psi$ functional is decomposed into nonlocal and local
parts, which are then approximated by {\it GW} and DMFT respectively.
This means that the local physics of all valence orbitals,
including rather itinerant $s$ or $p$ states, would be generated
from a self-consistent impurity model. 
It stands to reason
that the self-consistent dynamical $\mathcal{U}$ for those 
orbitals would in fact
come out to be rather small, so that the local dynamical 
contribution to the self-energy is also small and well described
by its first order term in $W$.
In practice, the self-energy for the itinerant states would 
thus be well described by a perturbative self-energy, that
is by the {\it GW} self-energy for both, local and nonlocal parts.

In view of these considerations, it seems a waste of computing
time to attempt to solve a dynamical impurity models for
all valence states, since the same result can be obtained
by applying the DMFT construction only to a subset of
``correlated'' states, and to treat all others entirely
by {\it GW}.
A scheme along these lines was proposed and implemented
in Ref.~\onlinecite{jmt_svo}.

The equations for the self-energy and polarization
are in this case replaced by
\begin{eqnarray}\label{Sig_a2}
\Sigma^{xc}(\vk,\iomn)_{LL'} &=& \Sigma_{GW}^{xc}(\vk,\iomn)_{LL'}
\\
&-& \sum_\vk \Sigma_{GW}^{xc,d}(\vk,\iomn)_{LL'}
+ [\Sigma^{xc,d}_{imp}(\iomn)]_{LL'}
\nonumber
\\
\label{P_a2}
P(\vq,\inun)_{\a\b} &=& P_{GW}(\vq,\inun)_{\a\b}
\\
&-& \sum_\vq P_{GW}^d(\vq,\inun)_{\a\b}+ P_{imp}^{d}(\inun)_{\a\b}
\nonumber
\end{eqnarray}
where the superscript $d$ denotes the projection onto
the low-energy correlated space.

One may be tempted to redefine the $\Psi$ functional
as the one of the {\it GW} approximation $GWG$, corrected for
its local part by DMFT only within the correlated subspace
(denoted here as $d$), as follows:
\begin{eqnarray}
\Psi[G,W] = GWG &-& G^{loc,d} W^{loc,d} G^{loc,d} 
\nonumber
\\
&+& \Psi_{imp}[G^{loc,d}, W^{loc,d}]
\end{eqnarray}
or, alternatively, by keeping the original decomposition into
local and nonlocal parts
\begin{eqnarray}
\Psi[G,W] = \Psi^{nonlocal} + \Psi^{local}
\end{eqnarray}
but approximating the local one by a combination of {\it GW} and DMFT
\begin{eqnarray}
\Psi^{loc}[G,W] = G^{loc} W^{loc} G^{loc} &-& G^{loc,d} W^{loc,d} G^{loc,d}
\nonumber
\\ 
&+& \Psi_{imp}[G^{loc,d}, W^{loc,d}].
\end{eqnarray}
Here, the superscript $loc$ denotes the projection on the
local component, and $d$ the projection onto the correlated
subspace.

Though appealing at first sight, such combinations
cannot be justified without further approximations on
a functional basis. This is due to the fact that screening
couples the correlated and itinerant subspaces, so that 
``downfolding'' of the interactions to obtain an effective
bare interaction within the correlated subspace necessarily
involves a decoupling approximation.
In the functionals above, this is born out of the 
difficulty of defining $W^{loc,d}$, as well as of 
postulating that $\Psi_{imp}$ is a functional of the
Green's function and screened Coulomb interaction of
the correlated subspace only.

Fortunately, in practice, these conceptual difficulties do not prevent
us from identifying a well-defined scheme, that corresponds
to a combination of {\it GW} and DMFT, where the impurity model
is used for the correlated space only. The application
to SrVO$_3$ presented below confirms the accuracy of such
a scheme.
In the following subsection we therefore describe
the ``orbital-separated scheme'' used in the present work,
where
only the local part of the self-energy of the \textquotedblleft
correlated\textquotedblright\ orbitals is calculated from the impurity model
and all other local and nonlocal components are
approximated by their first order expressions in $W$. 

\subsection{Orbital-separated scheme: the Equations}
\label{orbsepeqns} 

For the reasons discussed above,
in the orbital-separated scheme, one deviates
from the general prescription Eqs. (\ref{Sig_a}-\ref{P_a}) for the
self-energy and polarization by replacing their
local parts by their counterparts generated from
an impurity model within the correlated subspace
only. We outline in the following the iterative loop obtained
at the one-shot {\it GW} level but with full self-consistency
at the impurity level.
We call the correlated subspace $d$-space and
its complement the $r$-space.
Projections onto these spaces are noted by superscripts.
We furthermore assume that we dispose of a Wannier
basis which blockdiagonalizes the full LDA Hamiltonian,
and that the {\it GW} self-energy is block-diagonal in the
same basis.
The Wannier basis can be thought of as obtained from
the construction of maximally localized
Wannier functions in the $d$ and in the $r$
space separately. The assumption of a vanishing {\it GW} self-energy
block $\Sigma^{dr}$ in this basis is an additional approximation,
\cite{ferdi_down}
which is however very accurate, as we have explicitly
verified for our target compound SrVO$_3$.
We note that the common assumption in {\it GW} calculations
of a diagonal self-energy in the Kohn-Sham basis
is in fact a less justified approximation, and even this
is not a severe restriction for SrVO$_3$ \cite{PhysRevB.88.165119}.

Starting with a guess for the Weiss field and the
auxiliary Hubbard $\mathcal{U}$, the impurity model is
solved, that is the impurity Green's function $G_{imp}$
and screened Coulomb interaction $W_{imp}$ are obtained.
These are matrices in the orbital space of the correlated
states only. In order to obtain the full self-energy
and polarization the combined quantities
\begin{eqnarray}
\Sigma = \Sigma_{GW} - \Sigma_{GW}^{loc,d} + \Sigma_{imp}
\label{orb-sigma}
\\
P = GG - G^{loc,d}G^{loc,d} + P_{imp}
\label{orb-P}
\end{eqnarray}
involve ``upfolding'' to the full Hilbert space.

Then, the self-consistency equations for the determination
of the Weiss mean-field and the auxiliary dynamical $\mathcal{U}$
of the impurity model require the $d$ projections
$G^{loc,d}, W^{loc,d}$ of the local Green's function
and screened Coulomb interaction
\begin{eqnarray}
G^{loc} (\omega) = \sum_k [\omega + \mu - H_0 - \Sigma]^{-1}
\\
W^{loc} (\omega) = \sum_q [V_q - P]^{-1}
\end{eqnarray}
to equal their impurity model counterparts.
In the self-consistency cycle, they are used
to update the auxiliary impurity model quantities:
\begin{eqnarray}
\mathcal{G} = [ {G^{loc,d}}^{-1} + \Sigma_{imp} ]^{-1}
\\
\mathcal{U} = [ {W^{loc,d}}^{-1} + P_{imp} ]^{-1}
\label{orb-U}
\end{eqnarray}
The impurity model is solved for these new Weiss field
and dynamical $\mathcal{U}$, the resulting impurity
Green's function and screened Coulomb interaction are
obtained and the cycle is iterated
until self-consistency.

In the present work, we resort to a further simplification
allowing us to carry out the full self-consistency
cycle only for the one-body quantities (Green's function,
self-energy and Weiss field) in the correlated subspace, 
but to work with fixed
dynamical interaction $\mathcal{U}$.
This is achieved by approximating $P_{imp}$ in
Eq.~(\ref{orb-P}) non-selfconsistently by its RPA value
$G^{loc,d}G^{loc,d}$ 
leaving us, see Eq.~\ref{orb-P}, with $P=GG$ where the LDA Green's function
is used for $G$.
Furthermore, we replace Eq.~(\ref{orb-U}) by
\begin{eqnarray}
\mathcal{U} = \left[ \sum_q [ W^{-1} + P^d ]^{-1} \right]_d
\end{eqnarray}
projected on the $d$ space and its local component.
These approximations consist in taking as dynamical
impurity $\mathcal{U}$ simply the cRPA estimate
for the dynamical Hubbard interaction of the $d$ subspace. This is
done by partitioning the RPA
polarization into two contributions, $P^d$ and $P^r$,
calculated at the one-shot level from
the LDA electronic structure.
$P^d$ includes only $d-d$
transitions, while $P^r$ includes all the rest.

For SrVO$_3$, we consider the subset of t$_{2g}$
states as correlated, while oxygen $p$-, vanadium
e$_g$-, and strontium-$d$ states are considered
as $r$-space.
The scheme implemented in the present
work can then be summarized as follows:
\begin{itemize}
\item Obtain $\mathcal{U}(\omega)$ from a cRPA
calculation for a t$_{2g}$ low-energy subspace,
that is as matrix element 
\begin{eqnarray}
\mathcal{U}(\omega) =\langle 
{\mathrm t}_{2g} 
| \frac{V}{1-V(P-P^{t_{2g}})} | {\mathrm t}_{2g} \rangle 
\label{Eq:U-crpa}
\end{eqnarray} 
\item Obtain $\Sigma= GW$ from a one-shot {\it GW} calculation,
decompose it into the Fock part $\Sigma^x = G V$
and the correlation part $\Sigma^c = GW - GV$. 
\item Construct the one-body Hamiltonian
\begin{eqnarray}
H_0 = H_{\hbox{\tiny LDA}} - v_{\hbox{\tiny LDA}}^{xc} + \Sigma^x 
\label{H0}
\end{eqnarray}
where the LDA exchange-correlation potential has
been replaced by the Fock exchange $\Sigma^x$.
\item Construct an impurity model in the t$_{2g}$
subspace: start from an educated guess for the
Weiss field (in practice, at first iteration we
use the LDA local Green's function).
\item Solve the impurity for the Green's function,
that is calculate the expectation value
\begin{eqnarray}
G_{imp} (\tau) = - \langle \mathcal{T} c(\tau)
c^{\dagger}(0) \rangle_S
\label{Gimp}
\end{eqnarray}
using the impurity action
\begin{eqnarray}
S &=& - \int d \tau d \tau^{\prime}
c^{\dagger}(\tau) \mathcal{G} (\tau - \tau^{\prime})
c(\tau^{\prime}) 
\nonumber
\\
&+&  
\int d \tau H_{inst}
\nonumber
\\
&+& \int d \tau d \tau^{\prime}
\mathcal{\bar{U}}(\tau - \tau^{\prime})
n(\tau) n(\tau^{\prime})
\end{eqnarray}
Here, $H_{inst}$ denotes
the standard Hubbard-Kanamori Hamiltonian for t$_{2g}$
states, parametrized by the intra-orbital interaction
$U=\mathcal{U}(\omega=0)$, its inter-orbital counterpart
$U-2J$, and the inter-orbital interaction
for like-spin electrons
$U-3J$, which is reduced by the 
Hund's exchange coupling $J$.
The quantity $\mathcal{\bar{U}}(\tau - \tau^{\prime})
= \mathcal{U}(\tau - \tau^{\prime}) - 
U \delta(\tau - \tau^\prime)$
denotes the dynamical interaction without the
instantaneous part $U=\mathcal{U}(\omega=0)$.
\item 
From the impurity Green's function, obtain the
impurity self-energy via the Dyson equation
\begin{eqnarray}
\Sigma_{imp} = \mathcal{G}^{-1} - G_{imp}^{-1}
\end{eqnarray}
\item The full self-energy within the t$_{2g}$
space is obtained by combining the nonlocal
{\it GW} self-energy, projected onto this subspace,
with the impurity self-energy:
\begin{eqnarray}
\Sigma(k, i \omega) =   \Sigma_{GW} - \Sigma_{GW}^{loc,t_{2g}} +  \Sigma_{imp} 
\end{eqnarray}
\item Calculate the local Green's function within the
t$_{2g}$ space using the combined self-energy
\begin{eqnarray}
G^{loc}= \sum_k (i \omega + \mu - H_0 - \Sigma(k, i \omega))^{-1} 
\end{eqnarray}
\item and use this Green's function to update the Weiss
field:
\begin{eqnarray}
\mathcal{G} = [ {G^{loc,t_{2g}}}^{-1} + \Sigma_{imp} ]^{-1}
\end{eqnarray}
\item Go back to the solver step, that is calculate
the impurity Green's function (\ref{Gimp}) for the
impurity model defined by $\mathcal{U}(\omega)$
and the new Weiss field $\mathcal{G}$.
\item Iterate until self-consistency.
\end{itemize}

\subsection{The Bose Factor Ansatz}
\label{dala}

At the heart of the set of {\it GW}+DMFT equations is the solution of
an impurity model with dynamical interactions. As will be discussed
in the results section, the typical energy scale of variation of
the latter is the plasma energy, which for transition metal oxides
is an order of magnitude larger than the bandwidth. In this limit,
the solution of the dynamical impurity model can be greatly
simplified. Indeed, the Bose factor ansatz (BFA) within the ``dynamic
atomic limit approximation'' (DALA) introduced in 
Ref.~\onlinecite{PhysRevB.85.035115} yields an excellent approximation
to the full solution. In this scheme,
the Green's function of the dynamical impurity model is
obtained from a factorization ansatz
\begin{equation}
G(\tau )=
\left( \frac{G(\tau )}{G_{stat}(\tau )}\right) 
G_{stat}(\tau )
\sim
\left.\left( \frac{G(\tau )}{G_{stat}(\tau )}\right) \right|_{\Delta=0}
G_{stat}(\tau )
\label{factorization-approximation} 
\end{equation}%
where G$_{stat}$ is the Green's function for a static impurity
model with constant U=$%
\mathcal{U}(\omega =0)$, and the first factor is approximated by its
value for vanishing bath hybridization $\Delta$.\cite{PhysRevB.85.035115} 
The BFA yields an extremely efficient, yet accurate, way of 
solving the impurity model, as was checked by benchmarks against 
direct Monte Carlo calculations in
Ref.~\onlinecite{PhysRevB.85.035115}. 
It moreover allows for a transparent physical interpretation
of the arising spectral properties, since the spectral representation
(lower panel of figure 3) of the bosonic renormalization factor 
that enters equation (\ref{factorization-approximation}),
\begin{equation}
B(\tau)= \left.\left( \frac{G(\tau)}{G_{stat}(\tau)} \right) \right|_{\Delta=0}
\label{Eq:B}
\end{equation}
can be interpreted as the density of screening modes \cite{PhysRevB.85.035115}. 
The bosonic factor (\ref{Eq:B}) can be expressed in terms of the
frequency-dependent interaction as
\begin{equation}
B(\tau) =  \exp\left(- \int_0^{\infty} \frac{d \omega}{\pi} 
\frac{\Im\ \mathcal{U}
(\omega)}{\omega^2}
(K_{\tau}(\omega) - K_{\tau=0}(\omega))
\right)
\label{Eq:BU}
\end{equation}
with the bosonic kernel
\begin{equation}
K_{\tau}(\omega) =  
\frac{\exp(-\omega \tau) + \exp(- \omega(\beta - \tau))}{1 - \exp(- \omega\beta)}.
\label{Eq:K}
\end{equation}

\subsection{Technicalities}
\label{technical}

In the practical calculations for SrVO$_3$, we use the
experimental (perfectly cubic perovskite)
structure  
with lattice constant a=3.844\AA.
Calculations are performed at inverse temperature $\beta=10$ eV$^{-1}$
unless otherwise noted.
We perform a
maximally localized Wannier function 
construction\cite{PhysRevB.65.035109,RevModPhys.84.1419} 
for the t$_{2g}$ part of the Kohn-Sham spectrum within LDA.
A one-shot {\it GW} calculation is performed within the
full valence orbital space and then projected into
the t$_{2g}$ space.
The {\it GW} calculations are performed using a k-mesh of
8x8x8 k-points (4x4x4 for the ARPES spectra), 
which is then Wannier-interpolated
\cite{RevModPhys.84.1419}
to a dense grid of 27x27x27 k-points for the {\it GW}+DMFT
calculation.

The nonlocal self-energy is fixed at the one-shot level
from the initial {\it GW} calculation, and the frequency-dependent
interaction $\mathcal{U}(\omega)$ at its cRPA value as
discussed above.
At the DMFT level our calculations are fully self-consistent
for all one-particle quantities within the t$_{2g}$-space,
determining the self-consistent Weiss field that -- together
with $\mathcal{U}({\omega})$ -- defines the auxiliary impurity
model, self-consistently solved for fixed nonlocal-{\it GW} self-energies.
This loop is performed in imaginary
time/frequency space at an inverse temperature $\beta=10$ eV$^{-1}$, allowing at the same time for the
chemical potential to adjust self-consistently so as to
provide the correct particle number.
The resulting Green's functions
are analytically continued by means of a maximum entropy
algorithm, using the technology of Ref.~\onlinecite{PhysRevB.85.035115}
to access the high-energy features.

\section{Electronic Structure of SrVO$_3$}

Our target material, SrVO$_{3}$, has been the subject of intense
experimental and theoretical studies (for a review of
work until 1998 see \cite{Imada}).
In this section, we provide a brief 
summary of our previous knowledge about the
electronic properties of this material, in particular
concerning photoemission spectroscopy and the corresponding
theoretical works.

SrVO$_3$ crystallizes in the cubic perovskite structure:
the $V^{4+}$ ions are surrounded by oxygen octahedra,
and these octahedra occupy the sites of a simple cubic
lattice. The Sr$^{2+}$ cation sits in the center of the cubes.
The electron count leaves a single $d$ electron in the V-d
states, which is largely responsible for the electronic
properties of the compound.
The octahedral crystal field splits the V-d states
into a lower-lying threefold degenerate $t_{2g}$ manifold, 
thus filled with one electron per V, and an empty $e_{g}$ doublet. 
The compound exhibits a metallic resistivity with a Fermi liquid
$T^2$ behavior up to room temperature
\cite{Onoda1991281}
and temperature-independent Pauli paramagnetism without
any sign of magnetic ordering \cite{Inoue199761}. 
Hall data and NMR measurements confirm the picture of a
Fermi liquid with moderate correlations
\cite{Onoda1991281,Eisaki_phd}. 
These properties make SrVO$_3$ an ideal model material
for studying the effects of electronic Coulomb interactions.

\begin{figure}%
\includegraphics[width=0.55\columnwidth,angle=270]{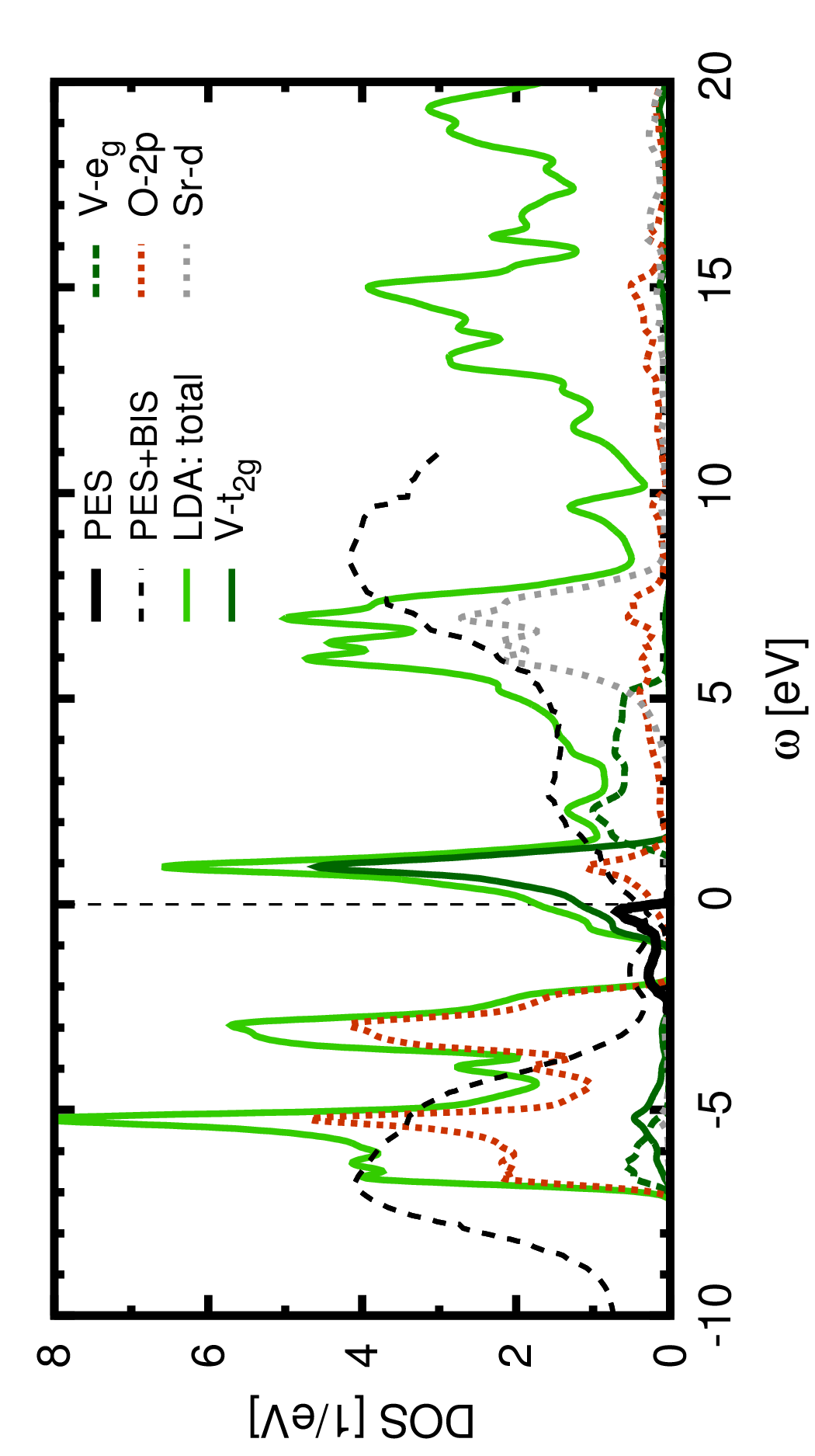}%
\caption{(Color online) 
Density of states within LDA in comparison with experimental spectra: PES\cite{PhysRevLett.93.156402} and PES+BIS\cite{PhysRevB.52.13711}.
} %
\label{fig:LDA}%
\end{figure}

Figure~\ref{fig:LDA} summarizes the Kohn-Sham electronic structure
of density functional theory within
the local density approximation (LDA):
the O-2p states disperse between -2 and -7 eV, separated from the
\t2g states whose bandwidth extends from -1 eV to 1.5 eV.
While the \t2g and e$_{g}$ bands are well separated at every given k-point,
the partial density of states (DOS) slightly overlap, 
and the e$_{g}$ states display a pronounced
peak at 2.3 eV. Finally, peaks stemming from the Sr-d states are
located at  6.1 eV and 7.1 eV.
We have superimposed to the LDA DOS the experimental PES and Bremsstrahl-Isochromat spectroscopy (BIS) 
curves taken from Refs.~\onlinecite{PhysRevLett.93.156402,PhysRevB.52.13711}. 
The comparison reveals the main
effects of electronic correlation in this material: as expected on 
quite general grounds, LDA locates the filled O-2p states at too high 
and the empty Sr-d manifold at too low energies.
The \t2g manifold undergoes a strong quasi-particle
renormalization with a concomitant shift of spectral weight,
both of which are effects beyond the one-particle
picture.
Photoemission studies \cite{PhysRevLett.69.1796} 
early on provided detailed information on 
the disagreement between the measured spectra
and the LDA DOS:
In the experimental spectra
the t$_{2g}$ spectral weight extends down
to binding energies of about -2eV, i.e.\ to 1eV lower than is found in LDA.
On the basis of comparison with the Mott insulating compound
YTiO$_3$ the observed additional peak between -1.5 eV and 
-2 eV was identified as a lower Hubbard band (LHB) -- 
due to the removal process of an electron from an 
atomic-like localized t$_{2g}$ state -- , whereas
the low-energy spectral weight was attributed to renormalized
but coherent band states.
A BIS study located 
an electron addition peak at energies around 2.7 eV
\cite{PhysRevB.52.13711}.

With the advent of dynamical mean field theory, explicit
calculations for spectra for an infinite-dimensional Hubbard model
became available \cite{bible}, supporting the idea of Hubbard
bands persisting in the metallic state.
The qualitative resemblance of the photoemission spectra
with the occupied part of the three-peak structure of the 
infinite-dimensional one-band Hubbard model suggested 
SrVO$_3$ to be a prototypical correlated metal, in which
the coexistence of quasi-particle states and Hubbard bands
as well as their dispersions could be studied.
Due to the high symmetry of the crystal structure, and
the resulting threefold degeneracy of the t$_{2g}$ bands,
it was moreover argued 
that a purely local self-energy would lead to ``pinning''
of the value of the fully interacting spectral function at 
the Fermi level to the one corresponding to the density
of states of the one-particle band structure. 
Any deviation from such ``pinning'' behavior\cite{pinning} can thus be taken as a proxy
for nonlocal components in the many-body self-energy
\cite{PhysRevB.52.13711}.

A difficulty arose from the extreme surface
sensitivity of the photoemission process, as evidenced
in Refs.~\onlinecite{Inoue_phd, PhysRevB.58.4372, 0295-5075-55-2-246, PhysRevB.73.052508}.
These authors performed systematic photoemission studies at
different photon energies, and witnessed
a pronounced
photon energy dependence of the quasi-particle peak,
which they rationalized as a varying surface sensitivity.
Measurements at high photon energies (900 eV)
\cite{PhysRevLett.93.156402} indeed found a more developed $t_{2g}$ quasi-particle peak,
in agreement with upcoming many-body calculations 
within dynamical mean field theory using the
LDA density of states (DOS) or the LDA Hamiltonian as input 
\cite{PhysRevB.72.155106, PhysRevLett.90.096401, pavarini:176403, 1367-2630-7-1-188}.
The increased intensity ratio of the quasi-particle and satellite feature thus suggested nonlocal self-energy
effects, neglected in DMFT, to be small.
Interestingly, even the surface sensitivity could
be modelled within such calculations
\cite{PhysRevB.73.245421}.
Angle-resolved photoemission spectra \cite{PhysRevLett.95.146404}
measuring the Fermi surface of SrVO$_3$ found
cylindrical Fermi sheets, in agreement with
theory, confirming the picture of a normal Fermi liquid.

Subsequent ARPES work adopted different 
strategies to increase bulk sensitivity: 
Laser ARPES \cite{PhysRevLett.96.076402} studied the very low-energy 
spectral features, finding a ``dip'' at the Fermi 
level or a maximum of the quasi-particle peak slightly
below (at around -0.2 eV). 
This work reopened the question about the role of nonlocal 
self-energy effects in the very low-energy 
properties of SrVO$_3$, since it remained unclear 
whether this feature is a result of the different 
experimental conditions of the laser ARPES setup
(restricted Brillouin zone sampling, matrix elements 
or other), or whether it reflects the true bulk
electronic structure at these very low energies.
For a half-filled one-band Hubbard model on
a cubic lattice, a similar ``dip'' effect was indeed
found within a cluster dynamical mean field
study \cite{PhysRevB.76.045108}.
Very recently, a realistic dynamical cluster
approximation study \cite{PhysRevB.85.165103} confirmed the
possibility of nonlocal effects inducing such
a depletion at the Fermi level.

Takizawa et al. used thin films with
atomically flat surfaces prepared
{\it in situ} \cite{PhysRevB.80.235104}, 
and were able to observe the band dispersions
not only of the coherent band but also of the
Hubbard bands. An interesting effect was
observed concerning the
lower Hubbard band: its intensity is
strongly momentum-dependent, with
its maximum in regions where also the
band states are occupied (k$<$ k$_F$),
whereas they fade away for k-points corresponding
to empty coherent bands \cite{PhysRevB.82.085119},
in agreement with theoretical modeling
within DMFT \cite{PhysRevB.80.235104}.
Recently, also SrVO$_3$-based hetero-structures have been
studied experimentally\cite{PhysRevLett.104.147601} and suggested
for electronic device applications\cite{2013arXiv1312.5989Z}.

The overall picture which emerges from all these works 
is that of a correlated metal with a
quasiparticle 
mass enhancement of about 2
\cite{PhysRevB.52.13711,PhysRevB.73.052508,
PhysRevLett.93.156402, PhysRevB.80.235104,
PhysRevLett.109.056401}
and a photoemission  (Hubbard--)satellite at 
around -1.6 eV binding energy.
This physics is reproduced by dynamical mean field 
calculations using the LDA electronic structure as input.
The first works 
\cite{PhysRevB.72.155106, PhysRevLett.90.096401,  
pavarini:176403,
1367-2630-7-1-188}
used a low-energy model comprising only the t$_{2g}$
manifold, where  
the local orbitals are constructed
from a downfolding procedure that incorporates also
the ligand O-2p tails. Different choices of such
orbitals were compared \cite{lechermann:125120},
demonstrating that as long as the considered energy
window is restricted to the t$_{2g}$ bands only,
results do not depend on the precise choice of the
local orbitals (maximally localized Wannier
functions, Nth order muffin tin orbitals, or projected
atomic orbitals).

SrVO$_3$ became the drosophila of combined LDA
and DMFT calculations, and new implementations were
quite systematically tested on this compound
(see e.g.\ \cite{lechermann:125120, 0953-8984-20-13-135227, amadon:205112, 
PhysRevB.80.085101, 0953-8984-23-8-085601}).
Apart from the effective t$_{2g}$ model, also
Hamiltonians including explicitly V-d and
O-2p ligand states in the non-interacting
Hamiltonian were used
\cite{amadon:205112, PhysRevB.80.085101, 0953-8984-23-8-085601}. 
It has been argued that the inclusion of ligand
states leads to more localized d-orbitals, and
an {\it a priori} better justification of the
local approximation made by DMFT.

Momentum-resolved spectral functions were 
calculated from dynamical mean field theory
in Ref.~\onlinecite{nekrasov:155112}, in agreement with
the experimental dispersion. They evidenced
an additional feature, a ``kink'' structure
at around -0.3 eV binding energy, which was
later on rationalized as a generally expected
phenomenon in correlated electron materials
\cite{byczuk-2007-3}: of purely electronic
origin, kinks appear at the crossover scale
at which the low-energy linear 
(Fermi liquid) behavior of the real part 
and the quadratic behavior of the imaginary
part of the self-energy cease to be valid.
In the meanwhile, kink structures observed
in other materials, e.g.\ LaNiO$_3$ \cite{PhysRevB.79.115122},
were also investigated theoretically
and have been consistently reproduced by
dynamical mean field calculations
\cite{PhysRevB.85.125137}.
For SrVO$_3$, the theoretical predictions
stimulated an intense search in photoemission spectra.
While Ref.~\onlinecite{PhysRevB.80.235104} still had to
conclude that ``the kink is weak and broad,
if it exists, but the curvature does indeed
change sign at around -0.2eV, as predicted'',
the very recent work by Aizaki et al. indeed 
identified such a kink
structure at around -0.3 eV \cite{PhysRevLett.109.056401}.

Besides dynamical mean field theory and
extensions, also other techniques of many-body
theory were employed to investigate SrVO$_3$.
A Gutzwiller study \cite{PhysRevB.79.075114}
investigated the mass renormalizations,
and renormalized densities of states as a
function of the Hubbard $U$.
Interestingly, to obtain the experimentally
observed mass enhancement a $U$ value beyond
5 eV was found to be necessary in this scheme.
Cluster model calculations systematically
addressed the spectroscopic properties of
SrVO$_3$ and analyzed the necessary ingredients
for a minimal model thereof  
\cite{mossanek:155127, mossanek:033104, PhysRevB.78.075103, JPSJ.78.094709}.
These studies emphasized the strong pd-hybridization, 
which is responsible for the large charge transfer
energy $\epsilon_d - \epsilon_p$.
Interestingly, an analysis of
the orbital character of the
different spectral contributions identifies
the spectral weight corresponding to the
t$_{2g}$ addition process as lying mainly between the
Fermi level and about 1 eV, in contradiction
with the dynamical mean field studies which
suggest an upper Hubbard band of t$_{2g}$
character at around 2.7 eV, that is at the 
precise location of the pronounced peak in
BIS spectra. The cluster model calculation
attributed this latter peak to the electron addition
into e$_g$ states \cite{mossanek:033104}.
We will come back to this point below.

With the advent of the constrained random
phase approximation (cRPA) \cite{PhysRevB.70.195104}
it became possible to calculate the values
of the local Coulomb interactions (``Hubbard
U'') specifically for the model under
consideration.
Again, SrVO$_3$ was chosen as a test material
to demonstrate the power of the method
\cite{PhysRevB.74.125106, PhysRevB.77.085122}, 
and it was shown that while U
values for a full model comprising ligand
states as well as V-d states can be as
large as 8 eV for the d-orbitals, for a
t$_{2g}$-only model the obtained value
was quite small: 3.5 eV.
The $U$ values used in the above cited
LDA+DMFT calculations, on the other
hand, varied rather between 4 eV
and 5.5 eV. These values were such
as to reproduce the observed mass
enhancement, even though the position
of the lower Hubbard band (LHB) was generally at slightly
too high binding energies, suggesting
that these values of $U$ were indeed on the
large side.
LDA+DMFT calculations with a $U$ value of 3.5 eV,
however, do not
reproduce the observed mass enhancement, nor
result in a clear LHB. 
This puzzle was solved only recently \cite{PhysRevB.85.035115}:
it was pointed out that
$U$ should be considered as a dynamical quantity
rather than a static interaction
\cite{PhysRevB.70.195104, PhysRevB.74.125106}. 
An LDA+$\mathcal{U}(\omega)$+DMFT
calculation taking not only the {\it ab initio}
value of the static component of $U=3.5$ eV but 
also its full frequency dependence into account
indeed reproduced the observed mass
enhancement as well as the position of the
lower Hubbard band \cite{PhysRevB.85.035115}.
This effect has very recently been
confirmed within an analogous study,
using a different impurity solver
scheme \cite{0295-5075-99-6-67003}. 

In the following, we briefly emphasize
a few puzzles, that remain within the
dynamical mean field description of 
SrVO$_3$, resulting from the above mentioned works.

\begin{itemize}
\item {\bf Inconsistency between LDA+DMFT
and cluster model calculations in the unoccupied
part of the spectra}
\\
While the assignment of orbital character
to the peaks in the spectral function made by
the cluster model calculations \cite{mossanek:033104}
coincides in the occupied part of the spectra with the
results of
dynamical mean field theory (or, to account also for 
the correct position of the LHB, of
LDA+$\mathcal{U}(\omega)$+DMFT), the position
of the upper Hubbard band (UHB) at 2.7 eV
found within the LDA+DMFT literature
is inconsistent with the cluster
model findings.
\item 
{\bf Interpretation of 2.7 eV BIS feature
as an upper Hubbard band inconsistent with
{\it ab initio} $U$ values}
\\
The interpretation of the BIS peak
at 2.7 eV as an UHB of t$_{2g}$ character,
done in the LDA+DMFT literature,
is inconsistent with the static
value of $U$ from cRPA. Indeed, from the
position of the LHB ($\sim$ -1.5 ev) and the
static $U$ value (3.5eV) one would expect
an UHB at 2 eV (as found in the 
LDA+$\mathcal{U}(\omega)$+DMFT
calculation \cite{PhysRevB.85.035115}). This leaves
the photoemission feature at 2.7 eV
unexplained within LDA+DMFT.
\item 
{\bf Position of O2p ligand states}
\\
LDA+DMFT calculations that also include
oxygen ligand orbitals, do not in principle
account for corrections to the LDA for these
orbitals. Such corrections have been introduced
by hand as an arbitrary shift on the O2p
states \cite{amadon:205112, PhysRevB.80.085101}.
This means that this position is not known
{\it ab initio} from LDA+DMFT.
On the other hand, it is well-known that
in the related compound SrTiO$_3$, which
is isostructural to SrVO$_3$ but of
d$^0$ configuration, the pd-gap of Kohn-Sham
theory within the LDA is underestimated
by 1.3 eV compared to experiment
\cite{benthem:6156}.
\item 
{\bf Position of Sr-4d states}
\\
An analogous problem arises when comparing
the energetic position of the Sr-4d
states in BIS and in Kohn-Sham
density functional theory, which underestimates
their energy by almost 2~eV.
By construction, combined LDA+DMFT schemes do not
correct for this error.
\item 
{\bf Relation between laser ARPES results
and nonlocal effects}
\\
To the best of our knowledge, it
remains open at this stage how to
reconcile the laser ARPES experiments
(and in particular the finding of a 
dip at the Fermi level) with the high-photon energy
PES which display a pronounced peak.
The study of nonlocal many-body effects
on a very low-energy scale remains thus
a challenging task for the future.
\end{itemize}

The present work addresses the first
four issues, leaving the last one
for future work.
In particular, we review and extend the
{\it GW}+DMFT calculations of Ref.~\onlinecite{jmt_svo}.
Since the publication of Ref.~\onlinecite{jmt_svo}
electronic structure calculations for SrVO$_3$ have met
renewed interest: besides a study \cite{PhysRevB.87.155147}
within the {\it GW} approximation (including a 
cumulant correction similar to the above
discussed Bose factor ansatz),
several groups have embarked into
attempts of setting up simplified schemes
mimicking the results of {\it GW}+DMFT%
\footnote{See e.g. Taranto et al. \cite{PhysRevB.88.165119}
for a study exploring the limits of an
implementation with static Hubbard interactions,
and -- most recently -- Sakuma et al.
\cite{PhysRevB.88.235110} who investigated the 
{\it ad hoc} combination of an
LDA+$\mathcal{U}(\omega)$+DMFT self-energy
with a {\it GW} one}.
Interestingly, while different elements
of the full calculations are indeed captured
in the different schemes, no scheme so far 
could fully reproduce the low-energy behavior,
and the question of designing approximate
schemes in a specific low-energy range
remains a largely open one. We will therefore
also devote an extended 
paragraph to a systematic comparison
of different approximate schemes and a 
discussion of what they can be expected to
provide.

\section{Results}

We now turn to the description of the results of
{\it GW}+DMFT calculations using the formalism outlined
above for our target compound, SrVO$_3$.
The {\it GW}+DMFT calculations will be put into perspective
by confronting them to pure {\it GW} calculations, as well
as to LDA+DMFT calculations both, with static and
dynamical interactions.
As a prelude, we discuss the dynamical Hubbard
interactions obtained for SrVO$_3$ within the
cRPA scheme.

\subsection{Dynamical interactions}

\begin{figure}
\includegraphics[width=\columnwidth,angle=270,trim= 10 10 100 20]{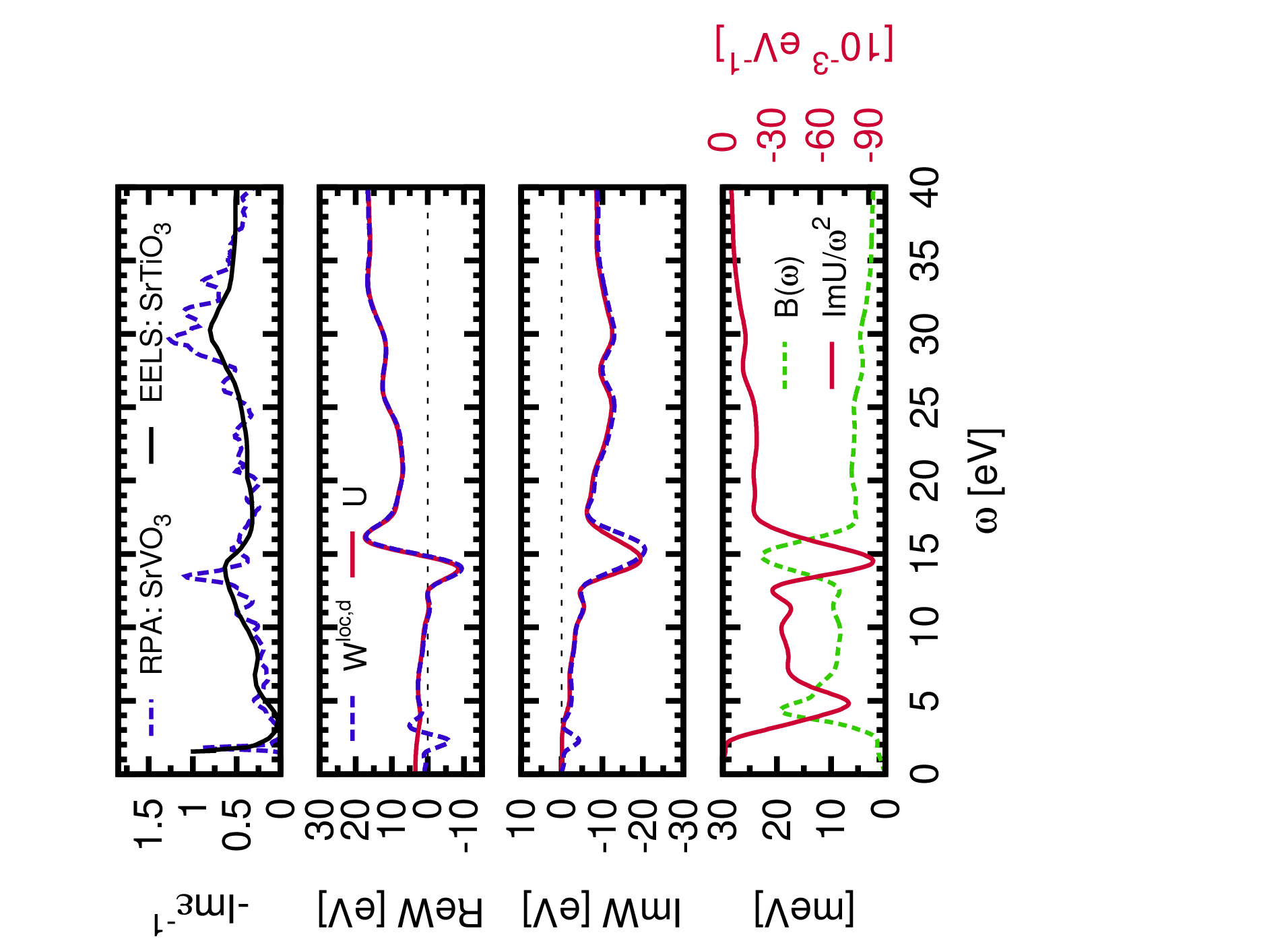}%
\caption{(Color online) 
Dynamical screening in SrVO$_3$. From top to bottom: (a) 
comparison of the inverse dielectric function of SrVO$_3$ within RPA with the experimental EELS spectrum of SrTiO$_3$\cite{PhysRevB.62.7964}.
(b)/(c)  real/imaginary par of the fully (partially) screened interaction $W^{loc,d}$ ($\mathcal{U}$) in the Wannier basis.
(d) Bosonic factor
(see text for definition) and density of screening modes $\Im\ \mathcal{U}(\omega)/\omega^2$.
} %
\label{fig:W}%
\end{figure}

In Figure \ref{fig:W}, we plot the screened and partially
screened Coulomb interactions:  $W$ denotes the matrix
element of the fully screened interaction in t$_{2g}$ 
maximally localized Wannier functions and the Hubbard $\mathcal{U}$ is defined in 
Eq.~(\ref{Eq:U-crpa}). 

The physical interpretation of the
frequency-dependence of the interactions
is transparent, if one recalls that the 
effective bare interaction within a subspace
of the original Hilbert space should include
screening by the omitted (e.g.~higher-energy- \cite{PhysRevB.70.195104} or
nonlocal-\cite{PhysRevB.86.085117}) degrees of freedom. 
Indeed, the net result of the rearrangement
of the high-energy degrees of freedom as response to
a perturbation of the system is an effective reduction 
of the perturbation strength in the low-energy space.
The effective Coulomb interaction
in a low-energy effective model for a correlated system
is therefore in general an order of magnitude smaller than the
matrix element of the bare Coulomb interaction.
Nevertheless, the latter is recovered in the 
limit of high-frequencies of the perturbation, when
screening becomes inefficient.
The crossover -- as a function of frequency -- from the 
low-energy screened regime to the high-frequency
bare matrix element of $\frac{e^2}{|{\bf r} - {\bf r}^{\prime}|}$
takes place at a characteristic screening (plasma) frequency 
where the dielectric function exhibits a pole structure.

For SrVO$_3$, the (partially) screened interaction, 
corresponding to the dynamical Hubbard interaction
at vanishing frequency, takes on a value of $U=3.5$ eV\cite{PhysRevB.77.085122} for the t$_{2g}$ orbital-subspace spanned by maximally localized Wannier functions.
The corresponding Hund's rule exchange $J$ is 0.6 eV.
The bare interaction, the matrix element of the
Coulomb interaction within the t$_{2g}$ Wannier
orbitals, equals $V=15$ eV. 
As seen in Fig.~\ref{fig:W}, the crossover from the low-energy
screened regime to the high-energy tail takes place 
at about 15 eV. At this energy, a well-defined plasma
excitation is observed. Indeed, 
the upper panel reproduces experimental electron energy loss (EELS) spectra
for the related compound SrTiO$_3$\cite{PhysRevB.62.7964}. This material is
isostructural to our target compound, and has one electron
less (d$^0$ configuration). The EELS data display a well-defined
plasmon excitation at about 15 eV.
The experimental spectrum is well-reproduced
by the theoretical imaginary part of the
inverse dielectric function calculated within the RPA.
The reason that, besides higher energy one-particle derived 
features, also the collective plasmon satellite of $d^0$ SrTiO$_3$
is well described by our calculation for the 
non-isoelectronic $d^1$ SrVO$_3$ resides in the fact that
it is not dominated by d-electron contributions. This is 
evident since the fully and partially screened interaction of the 
$t_{2g}$ orbitals, $W_{t2g}$ and $U_{t2g}$, are very similar at 
these energies.
Overall, this validates using the LDA electronic structure
for the purpose of calculating the effective interaction
$\mathcal{U}(\omega)$ of SrVO$_3$. 

The fully screened interaction $W$ furthermore exhibits a 
weaker feature at low energies ($\sim$ 2 eV), 
a ``subplasmon'', corresponding to 
a collective charge oscillation of the t$_{2g}$ charge
only. This peak is therefore not present when
the t$_{2g}$ screening processes are cut out, as is the
case in the construction of the effective interaction
$\mathcal{U}(\omega)$. As we will see later, this is the energy
regime where the local vertex corrections introduced by
DMFT modify the {\it GW} description of the spectral properties.
Features at these energies produced within {\it GW} calculations 
are thus not present any more in the {\it GW}+DMFT results
(see below).

In the many-body calculation, the frequency-dependent
interaction enters the bosonic factor $B(\tau)$ of Eq.~(\ref{Eq:BU})
in the form of $\Im\ \mathcal{U}(\omega)/\omega^2$.
This function can be interpreted as the density of
screening modes.
It is plotted in the lowest panel of Fig.~(\ref{fig:W}),
together with the spectral function of
$B(\tau)$ defined in Eq.~(\ref{Eq:B}).
Interestingly, these functions allow to identify yet another
feature, namely a well-defined peak at about 5 eV. We will
come back to this point later.

\subsection{{\it GW}}
\label{GW}

\begin{figure}%
\includegraphics[width=0.55\columnwidth,angle=270]{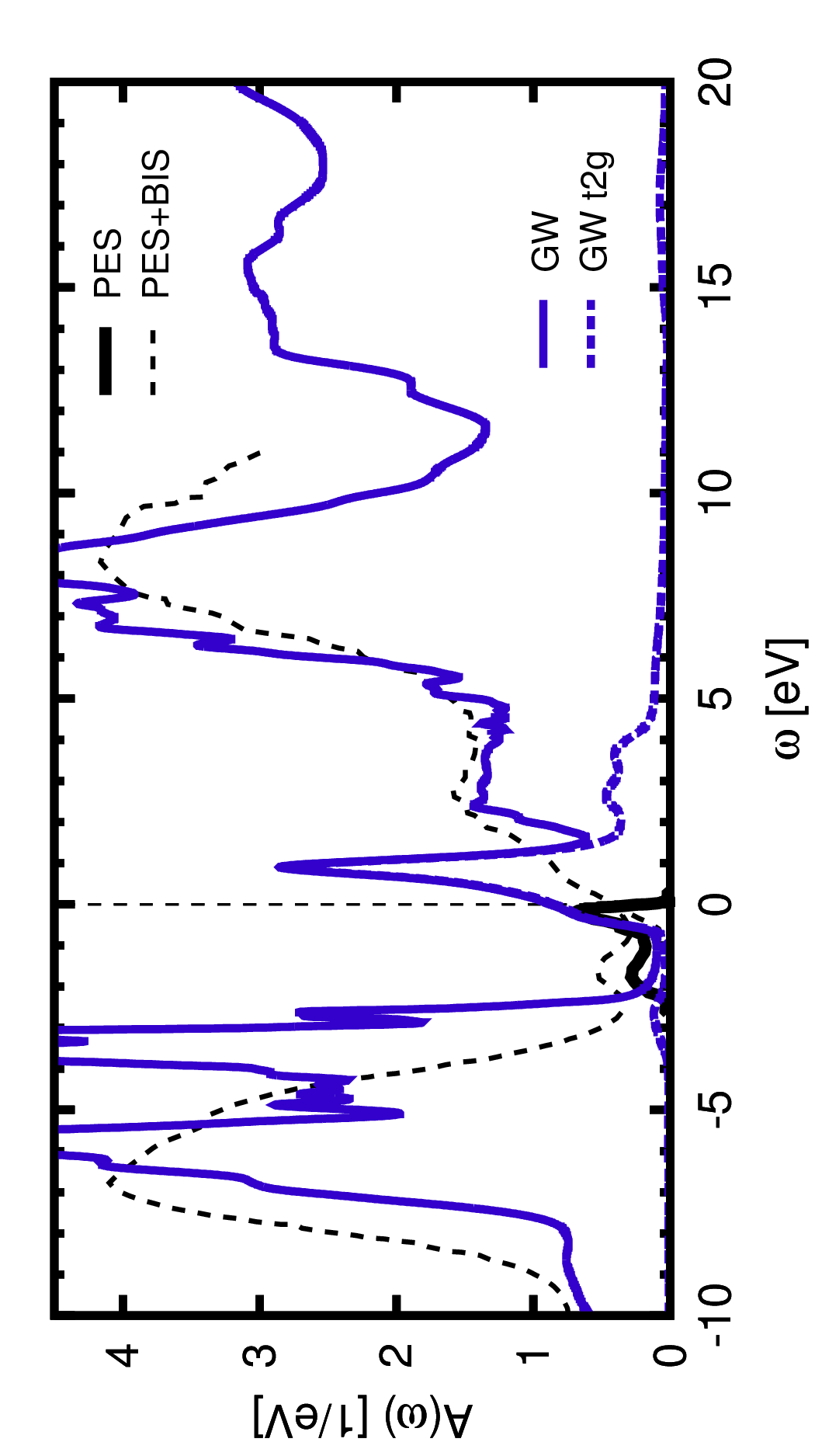}%
\caption{(Color online) 
The {\it GW} spectral function in comparison to the same experiments as in \fref{fig:LDA}.
} 
\label{fig:GW}%
\end{figure}

Several of the deficiencies of DFT calculations mentioned above can be addressed with Hedin's {\it GW} approximation\cite{hedin},
that uses the fully screened interaction $W$ discussed in the previous section.
We will in particular address the following two issues:

\begin{enumerate}
	\item {\bf higher energy states (O2p, Sr4d, ...).}
	Improvement of these is governed by exchange and correlation effects (beyond DFT)
	that (i) lie outside the realm of purely local interactions, and (ii) are beyond the (low energy / t$_{2g}$) orbital subspace.
	Thus inaccessible to DMFT-based methods, their correction is one pivotal merit that \GW\ contributes to 
	theories beyond DFT and DFT+DMFT.
	\item {\bf many-body effects at low energies.} 
Here we will discuss the impact of many-body renormalization on the t$_{2g}$ spectrum, with particular focus
on nonlocal self-energy effects (beyond DFT, and absent in DMFT).
\end{enumerate}

Besides a better description of the electronic structure of SrVO$_3$, our \GW\ calculation also gives useful fundamental
insights into the nature of correlation effects in transition metal oxides.
We will present evidence that dynamical and nonlocal correlation effects can essentially be separated (this was previously discovered for the iron pnictides and chalcogenides in Ref.~\onlinecite{jmt_pnict}). Further we will discuss the spatial extent of correlation effects in real space, 
putting into perspective corrections to the local picture of DMFT.

\subsubsection{Correction of higher energy features}

The {\it GW} spectral function is shown in \fref{fig:GW}. 
In the unoccupied part of the spectrum a substantial 
improvement over the LDA band-structure result, \fref{fig:LDA}, is seen:
states beyond the t$_{2g}$s are in excellent agreement with inverse photoemission results.
In particular the hump at around 2-5eV is very well captured. In contrast to assignments  in the DMFT literature, its spectral
weight stems largely from the vanadium $e_g$ states within \GW, in congruence with cluster based methods.\cite{mossanek:033104}
Beyond 5eV appear the Sr4d orbitals, again in remarkable accordance with the experimental intensity.

Also the position of occupied states, the O2p orbitals in the shown energy range, improve to the extent that
the experimental satellite at -1.6eV is no longer obscured by oxygen spectral weight.
With respect to the photoemission experiment however, the binding energy of the O2p is still too small by at least an electronvolt.
A possible remedy to this issue could be to extend the Wannier space to the O2p and vanadium $e_g$ states and include a local Hubbard interaction on the latter in the {\it GW}+DMFT.
This would favour a charge transfer into the O2p orbitals with which the $e_g$ states hybridize most, thus pushing the oxygen states further down.
In our {\it GW}+DMFT calculations here, we only consider the impact of local Hubbard interactions on the $t_{2g}$ subspace.

\subsubsection{Low energy renormalizations}

\begin{figure}%
\includegraphics[width=0.7\columnwidth,angle=-90]{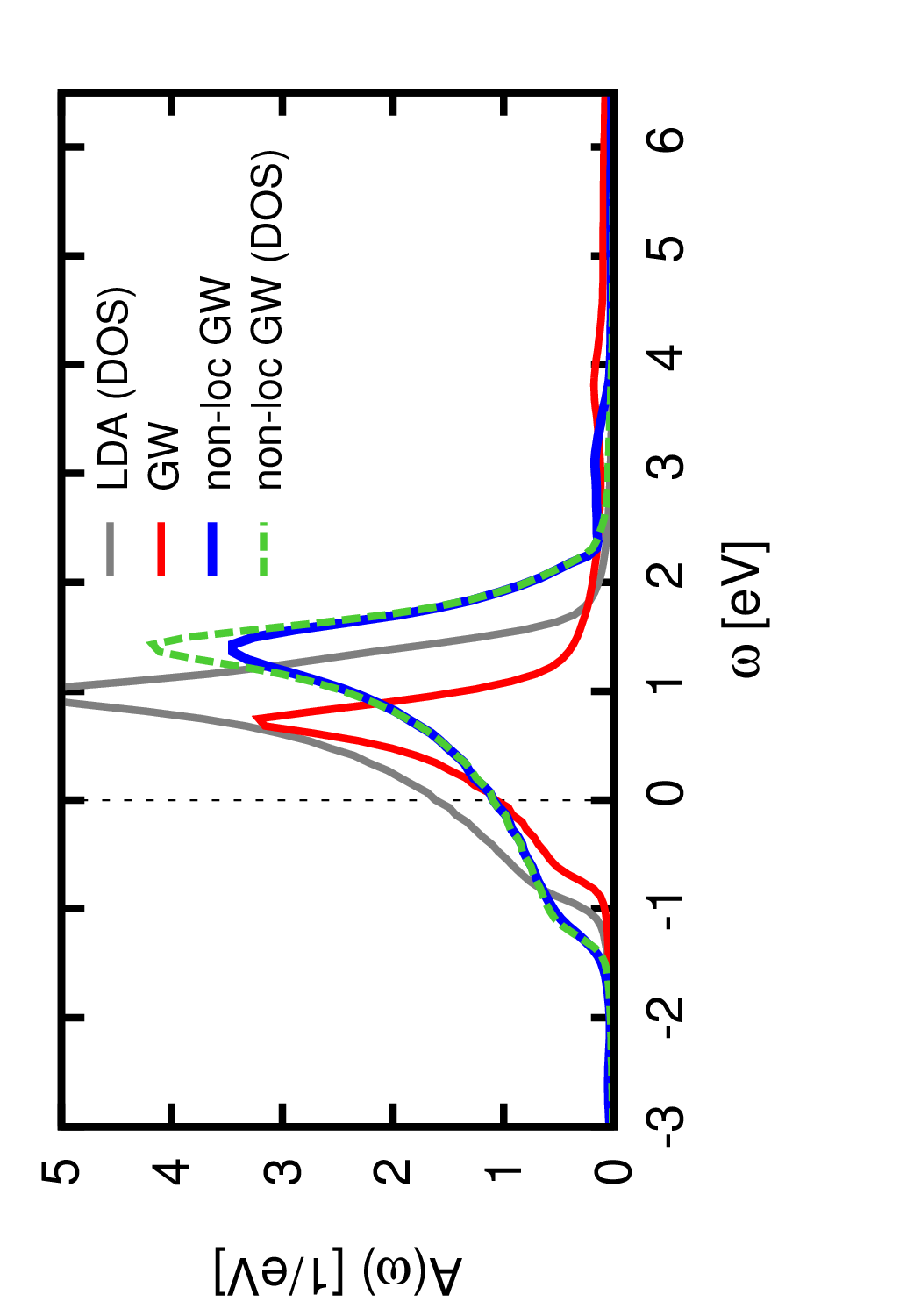}
\caption{(Color online) 
Comparison of the local {\it GW} spectral function with the LDA density of states.
Also shown is the spectral function and density of states of nonlocal-{\it GW}. See text for details.
} %
\label{awgw}%
\end{figure}
\begin{figure*}
\includegraphics[clip=true,trim=0 10 0 20,width=0.9\columnwidth,angle=0]{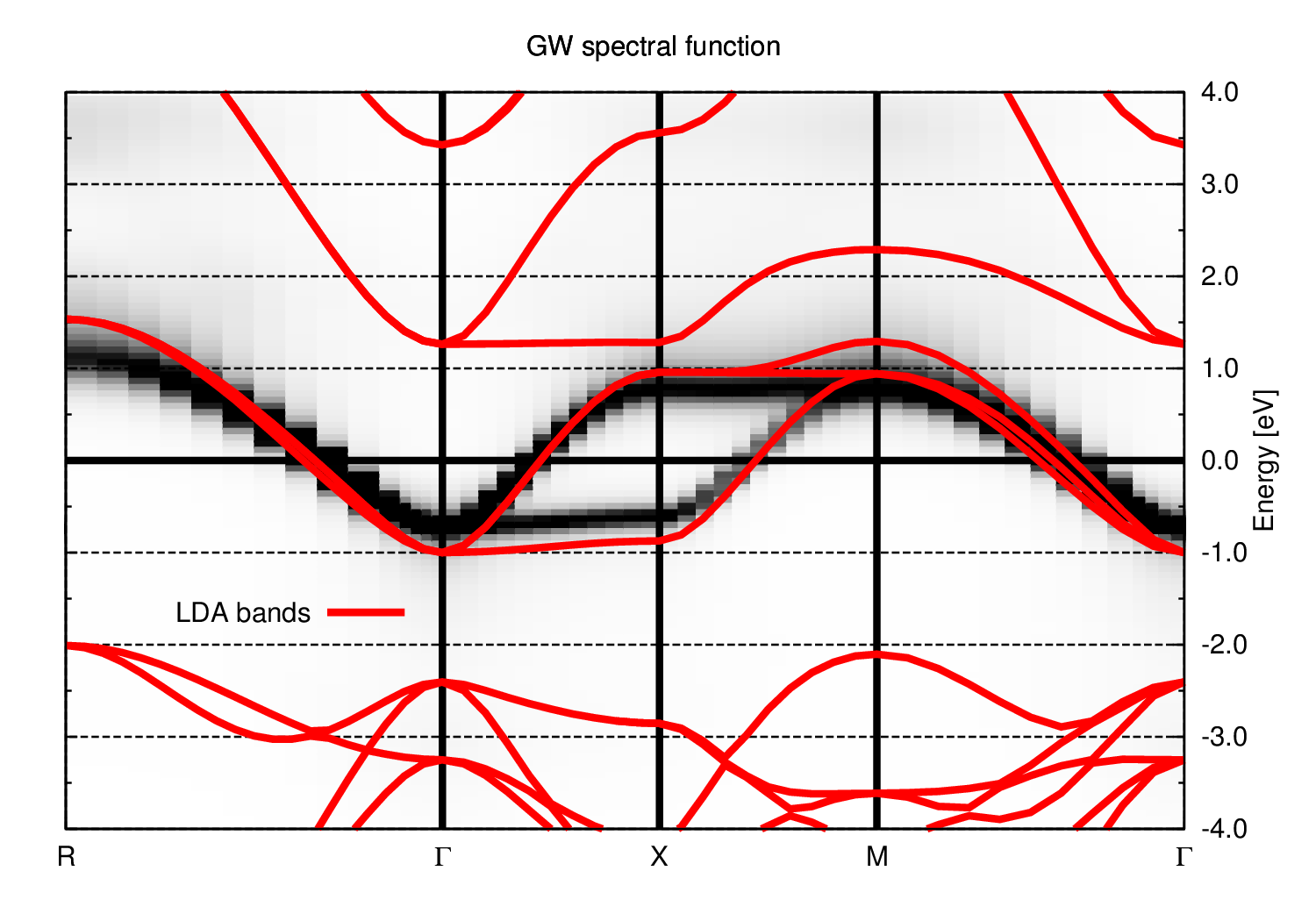}
\includegraphics[clip=true,trim=0 10 0 20,width=0.9\columnwidth,angle=0]{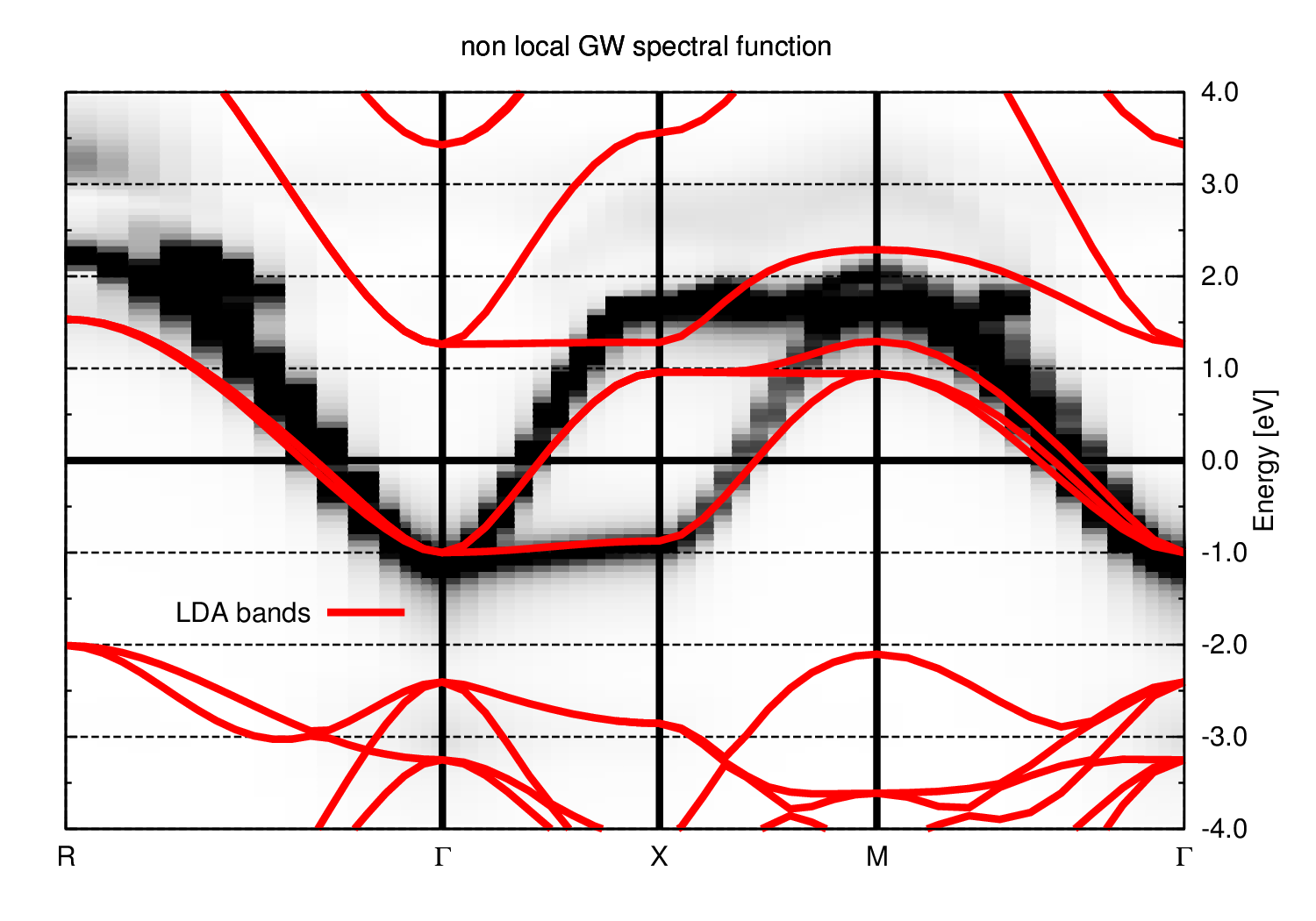}
\caption{(Color online) Momentum resolved spectral function 
(a) within the \GW\ approximation and 
(b) taking into account only the nonlocal part of the \GW\ self-energy.
Superimposed is the LDA band structure.
} %
\label{akwgw}%
\end{figure*}
Also shown in \fref{fig:GW} is the $t_{2g}$ contribution to the full spectral function.
The t$_{2g}$ bandwidth is reduced by about 25\% with respect to LDA, see also the momentum resolved spectra
in Fig.~\ref{akwgw} and Fig.~\ref{fig:arpes}.  
This suggests an overall effective mass $m_{\hbox{\tiny {\it GW}}}/m_{\hbox{\tiny LDA}}\sim1.3$.
The corresponding spectral weight is transferred to satellites that correspond to the features seen
in the fully screened interaction $W$, see \fref{fig:W}, namely at $\pm(\sim2)$eV as well as the $t_{2g}$ contributions to the plasmon satellite at 17eV.

To analyze the low-energy renormalizations further, we note that the mass enhancement 
relative to the LDA band masses is given by the ratio of the magnitudes of 
the group velocities within LDA, $\frac{d\epsilon_{\svek{k}i}}{dk_\alpha}=\jmbra{\Psi_{\svek{k}i}}   \partial_{k_\alpha} {H_{\hbox{\tiny {\hbox{\tiny LDA}}}}}(\svek{k})
\jmket{\Psi_{\svek{k}i}}_{k=k_F}$, and the {\it GW}
\begin{equation}
\frac{dE_{\svek{k}i}}{dk_\alpha}=\left.\frac{
\jmbra{\Psi_{\svek{k}i}}   \partial_{k_\alpha}\left( H_{\hbox{\tiny {\it LDA}}}(\svek{k})+\Re\Sigma_{\hbox{\tiny {\it GW}}}
(\svek{k},\omega)\right)
\jmket{\Psi_{\svek{k}i}} } 
{\left( 1- \jmbra{\Psi_{\svek{k}i}} \partial_\omega\Re\Sigma_{\hbox{\tiny {\it GW}}}(\svek{k},\omega) 
\jmket{\Psi_{\svek{k}i}}\right)}\right|_{k=k_F,\omega=0},
\label{landau}
\end{equation}
evaluated on the Fermi surface. 
Here, the self-energy is defined with respect to the LDA exchange-correlation potential: $\Sigma_{\hbox{\tiny {\it GW}}}=\Sigma^{xc}_{\hbox{\tiny {\it GW}}}-v^{xc}_{\hbox{\tiny LDA}}$.
Thus (besides a modified electron density), two ingredients for changes in effective many-body masses 
can be identified: 
(a) the {\it dynamical} part of the self-energy 
through the quasi particle weight $Z_\svek{k}=1/\left(1-
\partial_\omega\Re\Sigma_{\hbox{\tiny {\it GW}}}
(\svek{k},\omega)\right)_{\omega=0}$, and (b)
a renormalization via the {\it nonlocality} of the self-energy, $\partial_{k_\alpha}\Re\Sigma
(\svek{k},\omega)$.
In DMFT-based approaches, where the self-energy is local by construction, only the first mechanism
is present, hence $m_{\hbox{\tiny DMFT}}/m_{\hbox{\tiny LDA}}=1/Z_{\hbox{\tiny DMFT}}$.%
\footnote{
Of course, the LDA+DMFT self-energy will acquire a trivial momentum dependence when transformed from the local into the Kohn-Sham basis,
which is owing to the change in orbital characters for varying momenta.
}

The weight of the t$_{2g}$ quasi-particles in SrVO$_3$ is $Z_{k_F}\sim 0.53$ within the \GW\ approximation.
This is virtually the same value that is found for the homogeneous electron gas at the same density, $r_s=7.26$, when using the same method\cite{hedin}.
The low quasi-particle weight in conjunction with the only moderate bandwidth narrowing, $m_{\hbox{\tiny {\it GW}}}/m_{\hbox{\tiny LDA}}=1.3$, thus
advocates a notable enhancement of the group velocity, and thus band-width, from nonlocal correlations. 
We find it instructive to compute the spectral function when only taking into account these nonlocal effects.
To this effect, 
we take out the local part of the \GW\ correlation self-energy and construct 
$\Sigma_{GW}^{non-loc}(k,\omega)=\Sigma^{xc}_{GW}(k,\omega)-\left(\Sigma^{c \ loc}(\omega)-\Sigma^{c \ loc}(\omega=0)\right)-v^{xc}(k)$,
where
$\Sigma^{c \ loc}=\sum_k \Sigma^c_{GW}$.
The spectral function of this ``nonlocal-{\it GW}'' is shown in Fig.~\ref{akwgw}(b) for a selected k-path, while the local projection can be seen in Fig.~\ref{awgw}.
As anticipated, the t$_{2g}$ bandwidth is substantially widened. It becomes 44\% larger than the dispersion of the LDA.
In particular we see that this effect is notably more pronounced in the unoccupied part of the spectrum.
DFT being a theory to yield the correct ground state properties
(if the exact $v^{xc}$ was known), it seems natural that occupied states are 
better captured than unoccupied (excited) states (even though of course, the Kohn-Sham spectrum in principle has no physical meaning to begin with).
Also shown in Fig.~\ref{awgw} is the nonlocal-{\it GW} density of states, in which all (local and nonlocal) imaginary parts of the {\it GW} self-energy are omitted.
The presence of nonlocal correlation effects in the \GW\ approximation for SrVO$_3$ can also be evidenced as follows:
Indeed, for a purely local self-energy, and in the absence of orbital charge transfers (the t$_{2g}$-orbitals are locally degenerate), the value of the spectral function
at the Fermi level, $A(\omega=0$), is ``pinned'' to its non-interacting (LDA) value\cite{pinning}.
The violation of this pinning condition, see Fig.~\ref{awgw}, is thus heralding a nonlocal self-energy.
Obviously, the evidenced nonlocal renormalization is also beyond DFT+DMFT approaches, and hence another crucial contribution of the \GW\ to schemes such as \GW+DMFT.

\begin{figure}%
\includegraphics[width=0.8\columnwidth,angle=270]{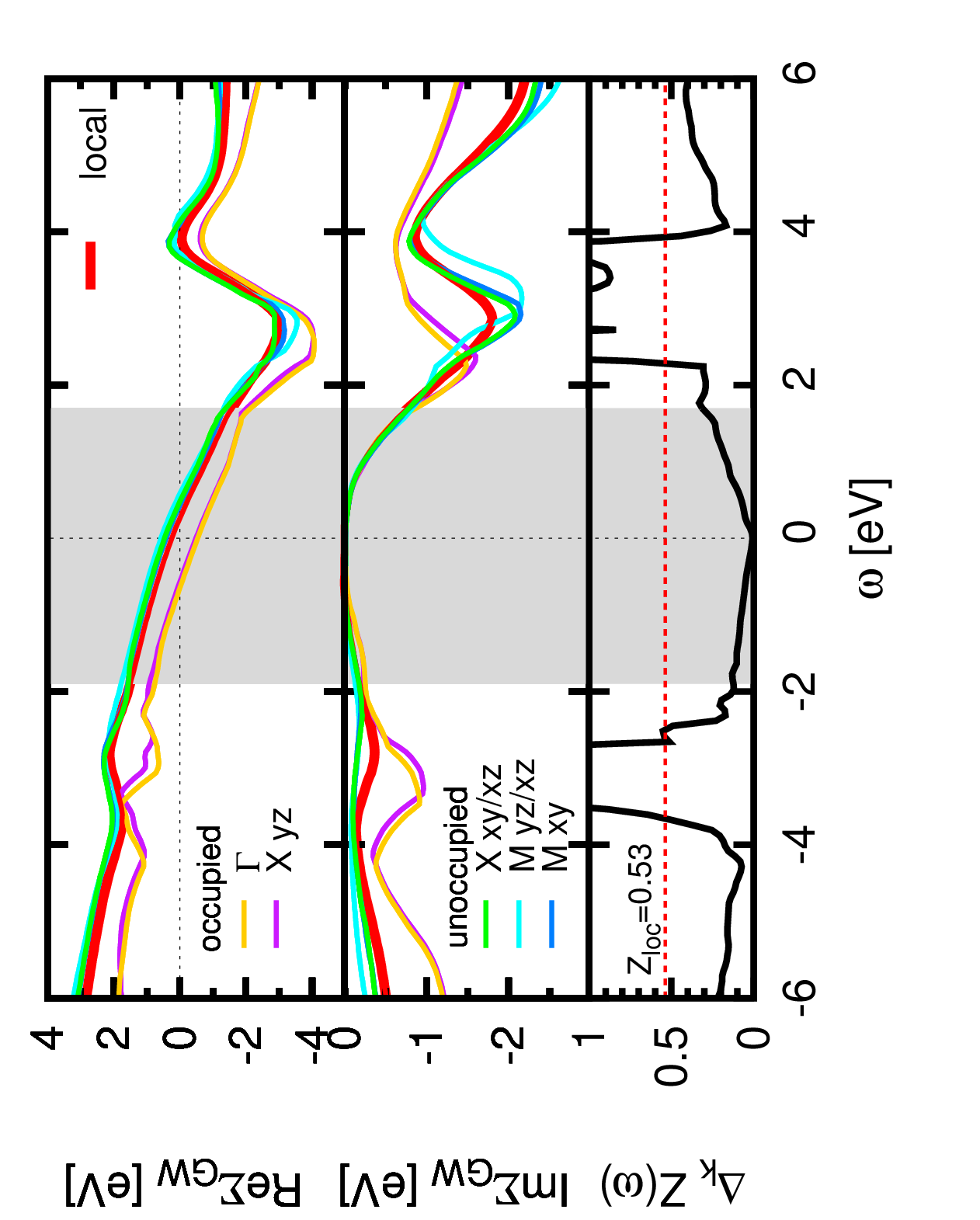}%
\caption{(Color online) 
The {\it GW} self-energy at several high symmetry points resolved into the three t$_{2g}$ (Wannier) orbitals as a function of frequency. 
Also shown is the local projection (real parts: top panel, imaginary parts: middle panel).
The lower panel displays the momentum variance of the frequency dependent generalization of the quasi-particle weight.
The origin of energy corresponds to the Fermi level and the shaded area roughly delimits the Fermi liquid regime within \GW.} %
\label{fig:Zw}%
\end{figure}

Thus, the fact that the LDA and full (local + nonlocal self-energy) \GW\ dispersion are somewhat comparable is owing to the competition 
of a band-width narrowing through the dynamics of the self-energy, and the tendency of nonlocal contributions to delocalize charge carriers.
However, the physics underlying these similar dispersions is very different: Indeed within the {\it GW}, almost half of the t$_{2g}$ spectral weight
is transferred to collective excitations at higher energies. This phenomenon is absent in effective one-particle theories such as DFT, but a 
physical reality,
see e.g.\ the EELS data in the preceding section.
However, due to the perturbative nature of {\it GW}, and its limitations regarding dynamical local correlations\cite{jmt_pnict},
it is not able to reproduce the lower Hubbard satellite seen in photoemission spectroscopy (\fref{fig:GW}).

\subsubsection{Separability of dynamical and nonlocal correlations}

Having discussed different ingredients to band-width renormalizations, we 
now examine the nature of correlation effects in more detail.

For the case of the iron pnictides and chalcogenides, Tomczak et al.~\cite{jmt_pnict} found
that -- within the {\it GW} approximation --
electronic correlation effects in the Fermi liquid regime are separable into
a dynamical self-energy that is local, and nonlocal contributions that are static.
This notion of locality holds when the self-energy is expressed in a local basis, in our case the maximally localized Wannier functions for the t$_{2g}$ subspace.
Does this empirical finding extend to the transition metal oxide SrVO$_3$?
In the upper panel of Fig.~\ref{fig:Zw} the real part of the \GW\ self-energy of SrVO$_3$ is shown for several high symmetry points in the Brillouin zone
as well as the local, i.e.\ momentum summed, element, as a function of frequency.
The offset, $\Sigma_{\hbox{\tiny \it GW}}(\omega=0)$, is positive for unoccupied orbital characters (xy/xz at the X point, and all t$_{2g}$'s at the M point, cf.\ Fig.~\ref{fig:arpes}),
and 
negative for the occupied orbitals. Thus (un)occupied spectral weight gets pushed (up) down in energy, 
congruent with the changes in the bandwidth seen in \fref{awgw} and \fref{akwgw}(b), as well as the reduction of the effective mass from
the value of the inverse quasi-particle weight $1/Z$. 

Regarding the frequency dependence, one can see that the self-energy is linear from roughly -2 to +1.8eV, which thus delimits the Fermi liquid regime within the \GW\ approximation.
The slope of the self-energy is slightly larger for $\omega>0$, thus compensating, in part, the static shift that is larger for unoccupied states.
Correspondingly, the imaginary part of the self-energy also grows faster with frequency in the unoccupied part, signalling stronger correlations for $\omega>0$.
The important finding here is that in the Fermi liquid regime, the frequency dependence (the linear slope in the real parts) at different momenta are very similar. 
That is to say that dynamical renormalizations in different regions of the Brillouin zone are comparable.
To investigate this more quantitatively,
we define
\begin{eqnarray}
Z_\svek{k} (\omega) = \left[ 1-\frac{\partial \Re\Sigma(\svek{k}, \omega)}{\partial \omega} \right]^{-1}
\label{Zk}
\end{eqnarray}
as a generalization of the quasi-particle weight $Z_{k_F}(\omega=0)$.
We further
introduce its momentum variance\cite{jmt_pnict} 
\begin{equation}
\Delta_k Z=\sqrt{\sum_\svek{k} Tr|Z_{\svek{k}}{(\omega)}-Z^{loc}(\omega)|^2}
\label{DZk}
\end{equation}
as defined with respect to the local projection $Z^{loc}$ of Eq.~\ref{Zk}, where the trace sums over the Wannier orbitals.
Then, $\Delta_k Z$ is a measure for the importance of {\it dynamical} self-energy effects that are {\it nonlocal}.
As is apparent from Fig.~\ref{fig:Zw}, $\Delta_k Z$ virtually vanishes at the Fermi level and is small compared to
$Z^{loc}(\omega=0)=0.53$ within the linear regime
\footnote{
It can be shown that the {\it linear} increase of $\Delta_kZ$ away from the Fermi level stems from the momentum dependence of $\mathcal{O}(\omega^2)$-corrections to $\Re\Sigma(\svek{k},\omega)$.
}.

This means that -- at least at the {\it GW} level\footnote{%
nonlocal correlation effects beyond the {\it GW} picture, stemming
e.g.\ from fluctuations in the spin-channel, are 
not included in this discussion. Also, the energy range of validity
of the Fermi liquid regime is generally overestimated within the
{\it GW} approximation, confining the argument to lower energies than
suggested by the {\it GW} picture.
} -- the dynamics of the quasi-particle renormalization is local, and, conversely, that nonlocal correlation
effects are static.\cite{jmt_pnict} As a consequence, the self-energy becomes {\it separable}:
The dynamical part is (almost) purely local, thus justifying the use of local but dynamic theories such as DMFT.
The nonlocal part, on the other hand, is static, as in theories employing generalized (orbital and momentum dependent) effective potentials such quasi-particle self-consistent (QS){\it GW}\cite{PhysRevLett.93.126406}%
$^,$\footnote{Naturally, a nonlocal dynamics is expected in lower dimensional systems, when spin fluctuations (not accounted for in {\it GW}) become
important, e.g.\ in the quasi-2d cuprates.
In Refs.~\onlinecite{ayral_gwdmft,PhysRevB.87.125149,PhysRevB.87.125149}, 
for example, it was found that nonlocal self-energy 
effects obtained from fluctuations in the charge channel are small
within \GW+DMFT calculations of an extended Hubbard model, indicating
that the leading nonlocal corrections are in the spin- rather than
the charge-channel.
}\
.

\begin{figure}%
\includegraphics[width=0.7\columnwidth,angle=270]{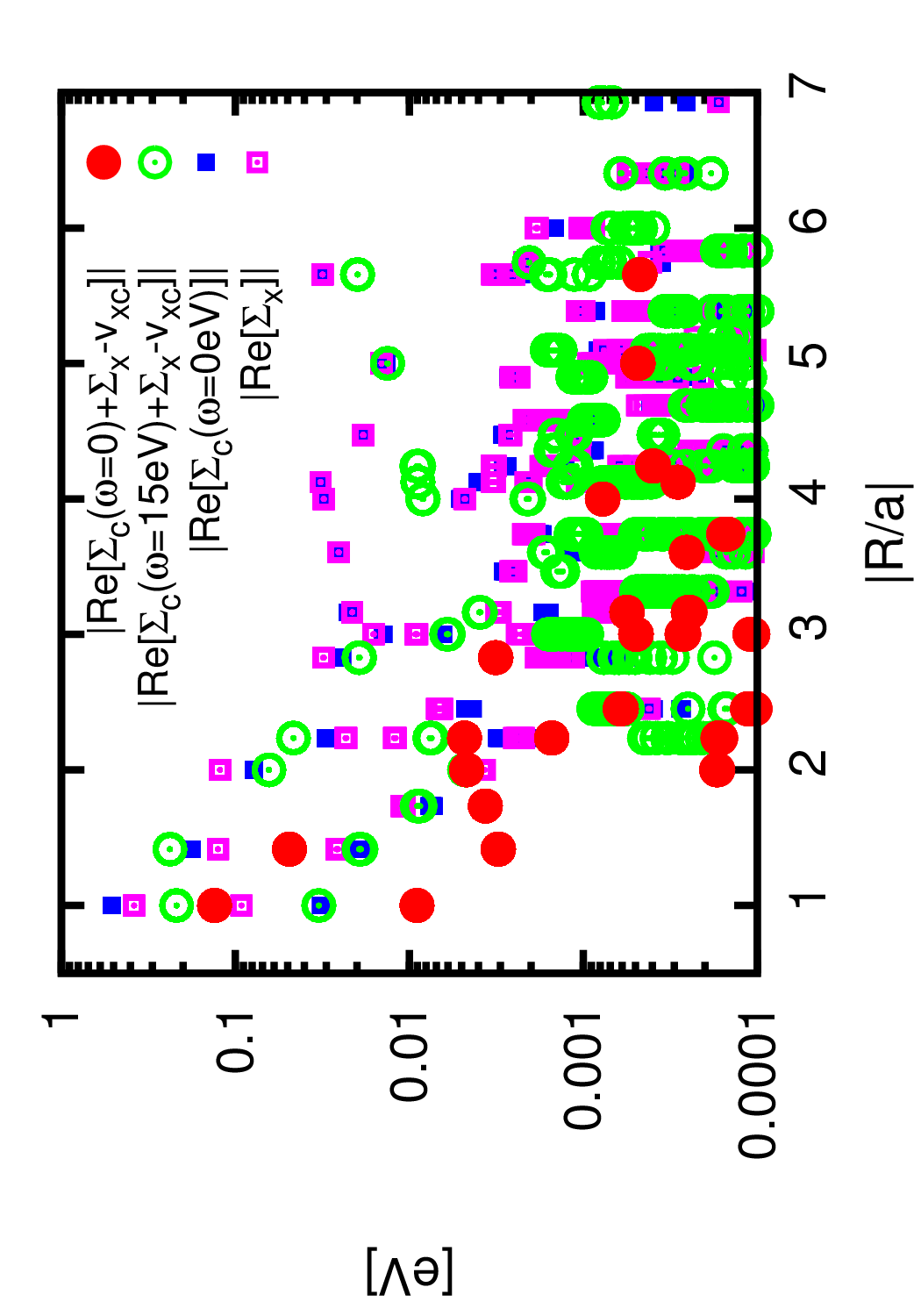}%
\caption{(Color online) The \GW\ self-energy correction to DFT as a function of real space distance at the Fermi level ($\omega=0$) and the energy region of the plasmon satellite ($\omega=15$eV).
The occurrence of several values per distance owes to different orbital orientations at growing numbers of neighbours.
Also shown are the absolute values of the ($\omega=0$) correlation and exchange self-energies of the {\it GW}.
} %
\label{Fig:Sr}%
\end{figure}

\begin{figure*}[t]%
\includegraphics[width=0.85\columnwidth,angle=0]{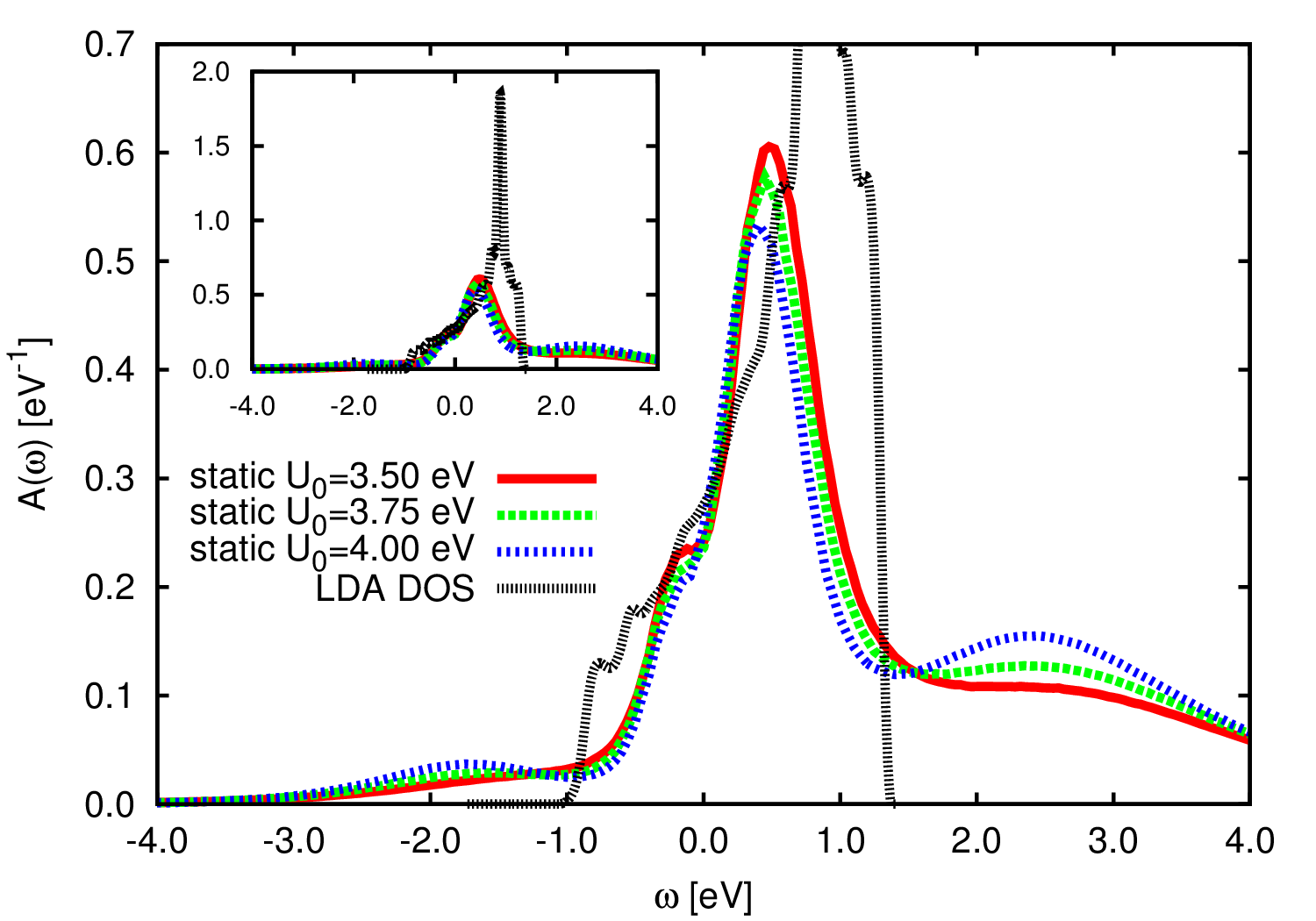}
\includegraphics[width=0.85\columnwidth,angle=0]{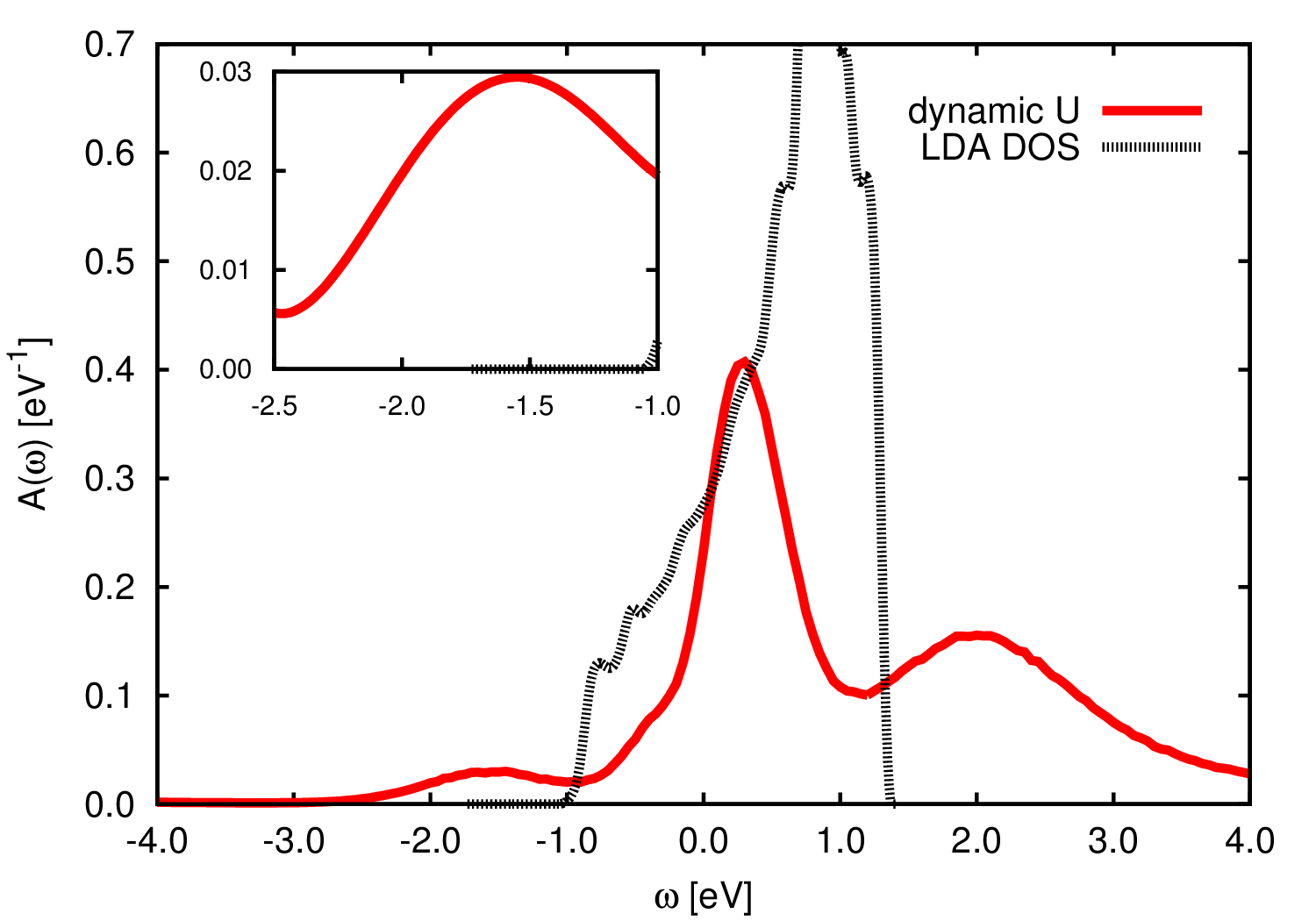}
\caption{(Color online) 
Local spectral function from standard LDA+DMFT with static
interactions (left).
LDA+DMFT with dynamical interactions (right).
Here, the spectral functions are normalized to one, such that the
filling corresponding to SrVO$_3$ is 1/6. 
} %
\label{Fig:LDA+DMFT}%
\end{figure*}

This non-trivial finding suggests that for many materials (in and for their Fermi liquid regime) the separation into local and nonlocal self-energies {\`a} la \GW+DMFT
simplifies to the extent that nonlocal correlations can be accounted for by a nonlocal yet static potential. 
This led the authors of Ref.~\onlinecite{jmt_pnict} to propose
a QS\GW+DMFT scheme, in which the QS\GW\ construction\cite{PhysRevLett.93.126406} is used to provide that potential.

\subsubsection{Bandwidth widening by nonlocal self-energy contributions}

We now discuss more in detail the widening of the band by nonlocal
self-energy contributions, as seen in Fig. \ref{akwgw}.
To this effect, we note that the separation of the self-energy
into a local dynamical and a nonlocal static part can 
be interpreted as a generalization of the familiar Coulomb-hole-screened
exchange (COHSEX) approximation to a full {\it GW} treatment.
Indeed, in the COHSEX approximation \cite{hedin} the 
{\it GW} self-energy is given by a static self-energy of the following form:
\begin{eqnarray}
\Sigma(r, r^{\prime}, \omega)
=
\Sigma_{SEX}(r, r^{\prime})
+
\Sigma_{COH}(r, r^{\prime})
\end{eqnarray}
where the first term is a screened exchange self-energy
built from the static screened Coulomb interaction
\begin{eqnarray}
\Sigma_{SEX}(r, r^{\prime}) = - \sum_{kn}^{occ}
\phi_{kn}(r) \phi^{\ast}_{kn}(r^{\prime})
W(r, r^{\prime}, \omega=0)
\end{eqnarray}
and the second contains the effect of the Coulomb hole
\begin{eqnarray}
\Sigma_{COH}(r, r^{\prime})
=
\frac{1}{2} \delta(r-r^{\prime}) (W(r, r^{\prime}, \omega=0) -
v(r - r^{\prime}))
\end{eqnarray}
Here, the indices $k,n$ denote Kohn-Sham states of wave
vector $k$, and the sum runs over occupied states only.
Interestingly, when separating the COHSEX self-energy
into local and nonlocal parts in the many-body sense
(that is, with respect to a localized basis set), the
nonlocal contribution stems from the screened exchange
self-energy only. For a system such as SrVO$_3$, the
{\it local} part of $\Sigma_{SEX}$ is -- by symmetry -- a 
scalar in the space of t$_{2g}$-orbitals, and can thus
be considered an irrelevant constant in that space.
The Coulomb hole self-energy, on the other hand, is
purely local.

The separation in static nonlocal and dynamical local
parts found in the preceding section can therefore be 
interpreted in the following way:
\\
(1) The nonlocal contribution to the self-energy can be
interpreted as a screened exchange self-energy 
$\Sigma_{SEX} = G W(\omega=0) - [G W(\omega=0)]_{local}$.
\\
(2) The local contribution contains the Coulomb hole
effect as well as band renormalizations beyond the COHSEX
approximation, stemming from the frequency-dependence of
the local dynamical self-energy.

Therefore, when considering the band structure corresponding to
the nonlocal self-energy contribution only, the Coulomb hole
part as well as the dynamical correlations are taken out 
since they are purely local, and the remaining 
correction can thus be interpreted as the screened exchange
contribution.
The widening of the band as compared to the Kohn-Sham band structure
is therefore the familiar broadening by exchange interactions (which,
here, are screened, thus leading to substantial but not as large
effects as in unscreened Hartree Fock theory).

The screened exchange self-energy correction to the
DFT exchange correlation potential can be written
as: 
\begin{eqnarray}
(\Sigma_{SEX} - v_{xc})(r, r^{\prime})
&=& - \sum^{occ}_{k^{\prime}n^{\prime}}
\psi^{\ast}_{k^{\prime}n^{\prime}}(r) \psi_{k^{\prime}n^{\prime}}(r^{\prime})
\\
&\times&\left(W(r, r^{\prime}, \omega=0)
- \delta(r - r^{\prime}) \tilde{v}(r)\right)
\nonumber
\end{eqnarray}
with a potential $\tilde{v}(r)$ representing the 
Kohn-Sham exchange-correlation contribution.

Matrix elements of this quantity in the Kohn-Sham basis
read
\begin{eqnarray}
&&
\langle k_0 n_0 | \Sigma_{SEX} - v_{xc} | k_0 n_0 \rangle
\nonumber
\\
&=& - \sum^{occ}_{k^{\prime}n^{\prime}}
\int d^3r \int d^3r^{\prime} 
\psi^{\ast}_{k^{\prime}n^{\prime}}(r) \psi_{k^{\prime}n^{\prime}}(r^\prime)
W(r, r^{\prime}, \omega=0)
\nonumber
\\
&&\times\psi^{\ast}_{k_0 n_0}(r) \psi_{k_0 n_0}(r^\prime)
+\int d^3r \tilde{v}(r)
n(r) |\psi_{k_0 n_0} (r)|^2
\end{eqnarray}

An intuitive inspection of these matrix elements suggests
the resulting correction to be small for occupied $|k_0n_0\rangle$ 
states, but to result in an upward shift for unoccupied states.
Indeed, for unoccupied states, the matrix elements
$\langle k_0 n_0 k^{\prime}n^{\prime}| W | k_0 n_0 k^{\prime}n^{\prime}\rangle$
are necessarily between product states that mix occupied
$|k_0n_0\rangle$ and unoccupied states $|k^{\prime}n^{\prime} \rangle$,
and thus small compared to $V_{xc}$.
This results in the familiar effect of a {\it GW} correction to 
conduction band states in simple semiconductors,
leading to a ``scissors'' correction to the too small 
Kohn-Sham band gaps.

In the case of the metallic SrVO$_3$ with d$^1$ filling, 
the band widening by
nonlocal contributions is much stronger for the unoccupied
part of the spectrum (which is enhanced by more than 1 eV)
than for the occupied part.
As we will see below, this effect will carry through the
\GW+DMFT treatment, where the screened exchange band structure
becomes renormalized by local dynamical correlations encoded
in the DMFT self-energy.

\subsubsection{The spatial range of correlations}

Having established the importance of nonlocal 
correlation effects, as well as their static nature at low energies, we want to characterize their 
extent
{\it in real space}.
Indeed there are efforts to extend DMFT calculations from the single impurity setup to a cluster of several sites (or several momenta) even for {\it ab initio} calculations.
For the case of SrVO$_3$ this was first done in Ref.~\onlinecite{PhysRevB.85.165103} using the dynamical cluster approximation (DCA) method, that 
partitions the Brillouin zone into momentum patches (two patches, in the cited work) and thus gives momentum resolved information on a coarse grid.

Here, we will rather follow the spirit of cellular DMFT, in which real-space clusters are embedded into the solid, 
thus allowing nonlocal correlations of the range of the cluster size.
The important question now is how big that cluster has to be in order to exhaust the 
extent of pertinent nonlocal correlations.
For this we note that self-energy diagrams beyond \GW\ give mainly local contributions\cite{PhysRevLett.96.226403}, and thus our findings based on the \GW\ approximation
are expected to have a wide range of validity\footnote{See, however, the preceding footnote.}. 

In Fig.~\ref{Fig:Sr} we show the magnitudes of the \GW\ self-energy corrections with respect to LDA at the Fermi level ($\omega=0$) and at energies near the plasmon peak ($\omega=15$eV)
as a function of the real space distance to a reference vanadium atom.

At the Fermi level, this correction is indeed rather short-ranged: Already at the next-nearest (vanadium) neighbour it has decreased by one order of magnitude.
This advocates that a 2x2x2 unit-cell cluster (beyond current computational capabilities) might already give meaningful results.
In the region of the collective (plasmon) excitation, the decrease in magnitude occurs more slowly, suggesting much larger cluster sizes.
This does not come as a surprise, since at these energies collective
long-ranged excitations are dominant.

\subsection{DMFT}
\subsubsection{DMFT with static interaction}

SrVO$_3$ has been used as a benchmark compound for standard
LDA+DMFT calculations, both within a low-energy description
comprising only the t$_{2g}$ states 
\cite{}
and including the oxygen ligands 
\cite{}.
It was argued that the static Hubbard interactions have to
be at least as large as 4 eV to reproduce the experimentally
observed mass enhancement.
The local spectral function then displays a three-peak
structure as in the correlated metal phase of the half-filled
single-band Hubbard model, even though the low filling of
1 electron in 3 bands makes the spectra highly asymmetric.
The lower Hubbard band, at $U$=4 eV, is located at slightly
too low binding energy (nearly -2eV, instead of the experimentally
observed -1.5 eV).
At about 2.5 to 2.7 eV, an upper Hubbard band is found.
Since this feature coincides in energy with an experimentally
observed electron addition peak, the LDA+DMFT literature has
thus identified the latter as an upper Hubbard band (see
however the \GW\ spectrum in \fref{fig:GW} and the discussion below).
When using the static component of the Hubbard interaction
calculated within cRPA ($\sim$ 3.5 eV), however, a very weakly
correlated metal is obtained, where the lower Hubbard band
is barely a shoulder structure and the mass enhancement
is much smaller than the experimentally observed one.
Figure \ref{Fig:LDA+DMFT}(a) reproduces the 
local spectral function for $U$ values varying between
3.5 eV and 4 eV, as calculated in Ref.~\onlinecite{lechermann:125120}.

\subsubsection{DMFT with dynamical interaction}

\begin{figure*}[t!]%
\includegraphics[width=0.55\columnwidth,angle=270]{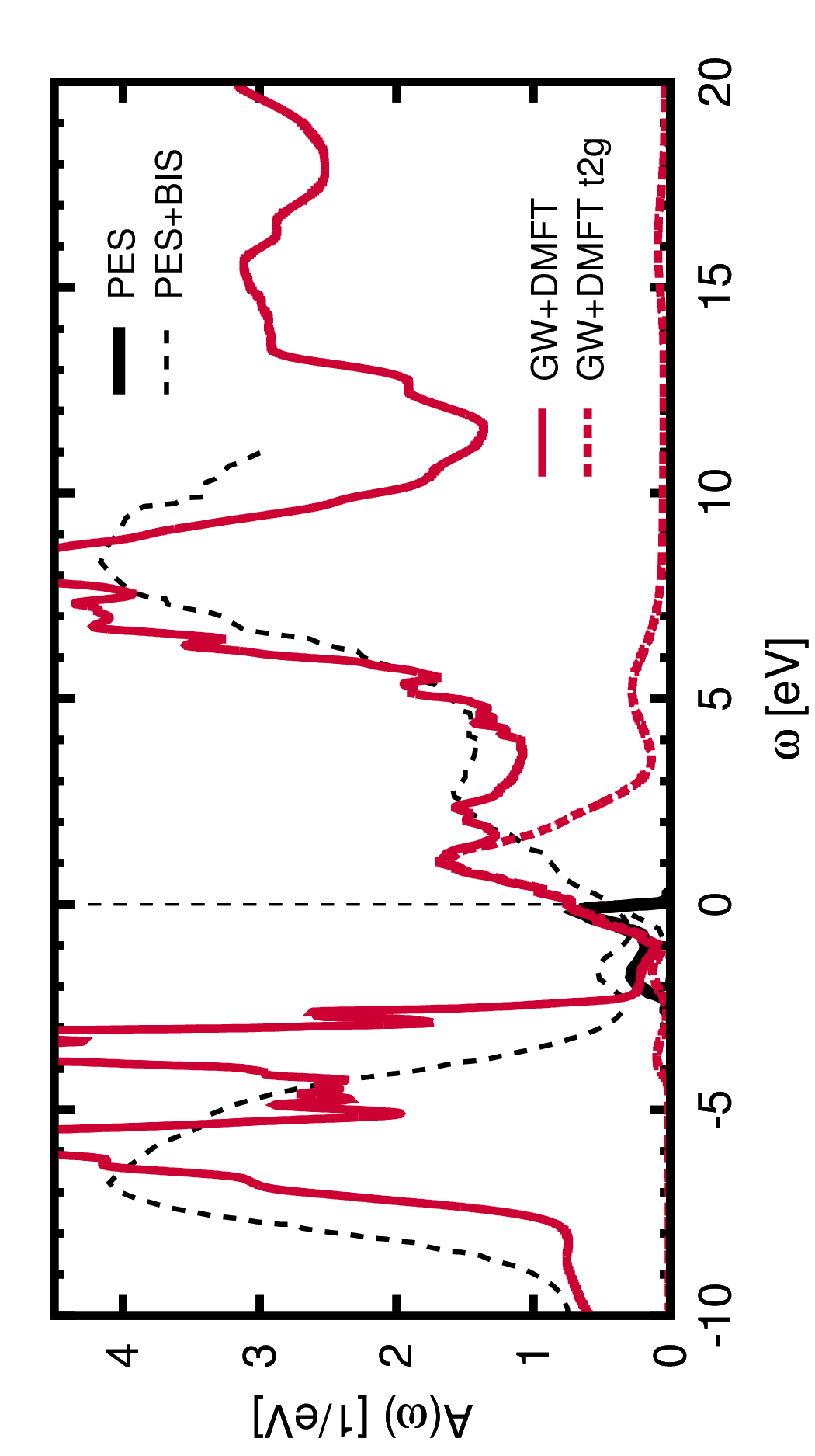}%
\includegraphics[width=0.55\columnwidth,angle=270]{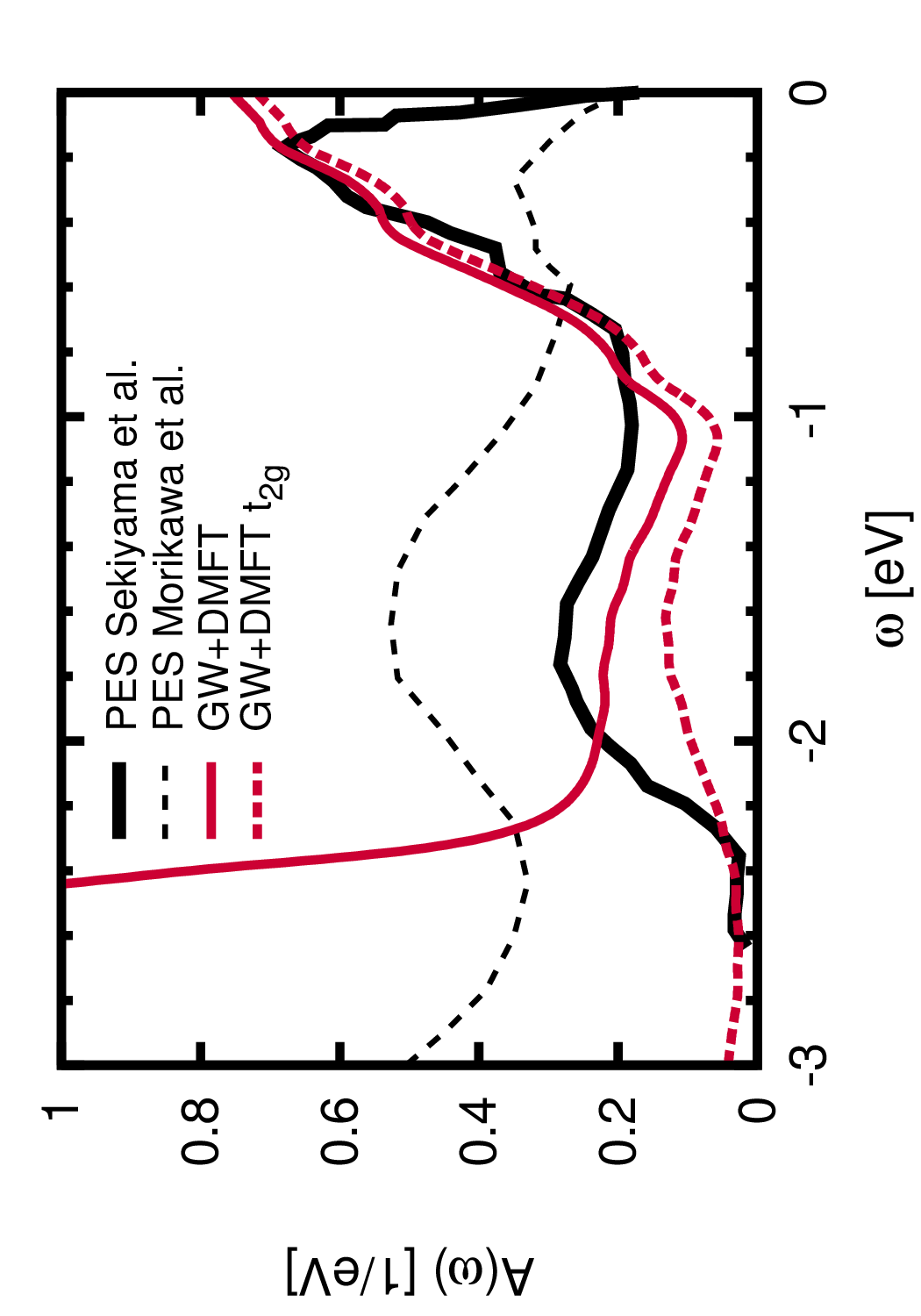}%
\caption{(Color online) 
{\it GW}+DMFT spectral function, in comparison to (inverse) photoemission spectra (same experiments as in \fref{fig:LDA}).
} %
\label{GWDMFT}%
\end{figure*}

The puzzle of the too weak mass renormalizations within LDA+DMFT
when the static component of the cRPA $U$ is used was solved
when it was realized that taking into account the frequency-dependence
of the interactions leads to additional mass enhancements \cite{PhysRevB.85.035115}.
Indeed, the high-energy tail of the dynamical interaction
{\it alone} was shown to be at the origin of a mass enhancement of
$Z_B^{-1}$ with $Z_B=0.7$ \cite{PhysRevB.85.035115}. The overall mass
enhancement of the calculation with the dynamical cRPA interaction
is $m^*/m_{\hbox{\tiny LDA}}\sim 2$, in reasonable agreement with ARPES
estimates.
Since, however, the static component of $U$ is smaller than what was
used before in static LDA+DMFT calculations, the position of the
lower Hubbard band is shifted towards the Fermi level,
correcting the deficiency of LDA+DMFT discussed above.
On the unoccupied side of the spectrum, an upper Hubbard band
feature appears at about 2 eV, substantially lower than what
was discussed within LDA+DMFT. Experimentally, such a feature
is not clearly resolved.
We can thus summarise the effect of dynamical interactions
within LDA+$\mathcal{U}(\omega)$+DMFT calculations by noting
that the only notable modification in the electronic structure 
is the improved description of the lower Hubbard
band, compared to experiment, whereas the situation is less
clear for the unoccupied part of the spectrum. We will argue
below that this scheme is actually as little appropriate
for unoccupied states as is the standard static LDA+DMFT.

\subsection{{\it GW}+DMFT}

\subsubsection{Full calculations}

\begin{figure}%
\includegraphics[width=\columnwidth,angle=270]{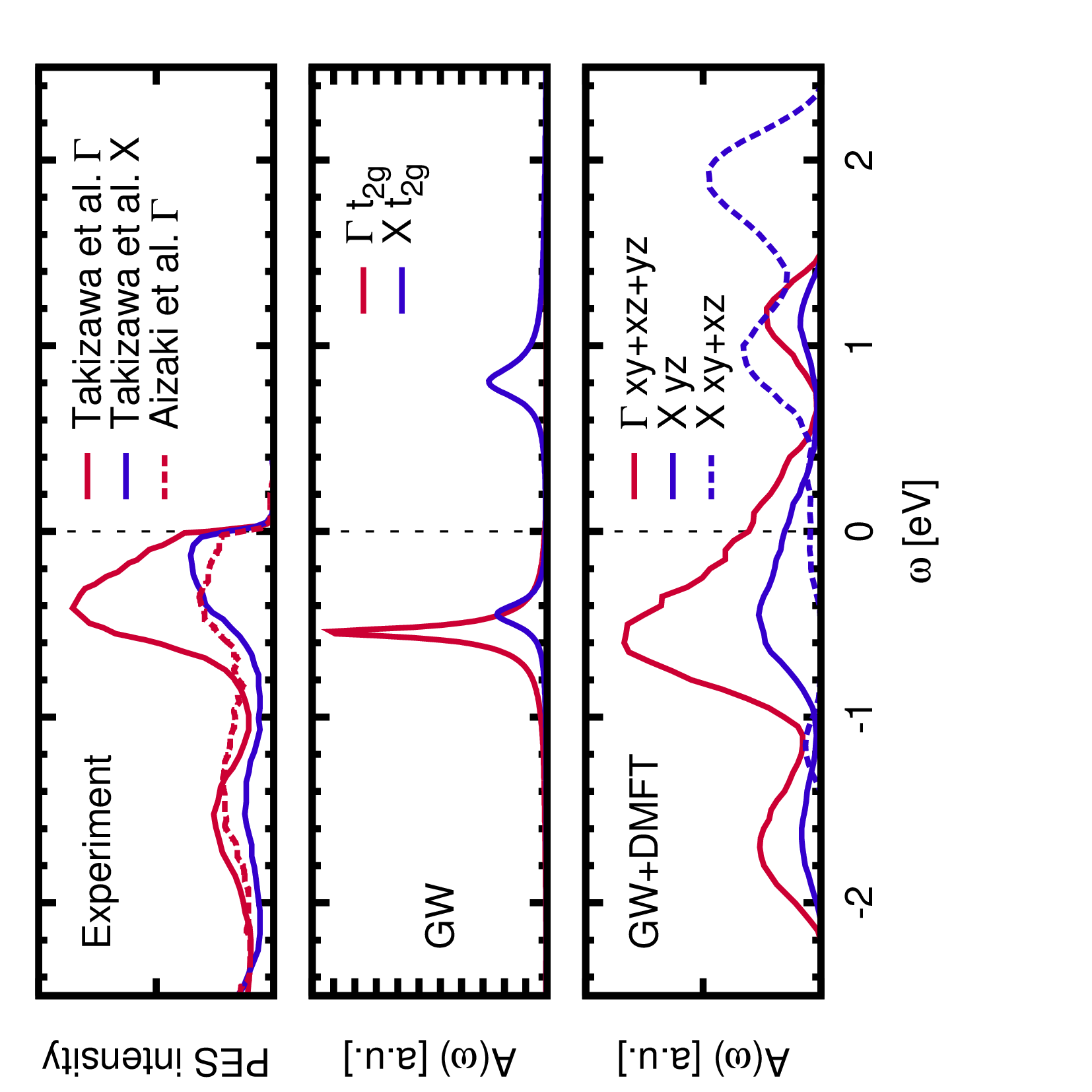}
\caption{(Color online) 
Momentum-resolved spectral function from {\it GW} and {\it GW}+DMFT  in comparison to photoemission experiments.
} %
\label{fig:arpes}%
\end{figure}

We now discuss the results of our combined {\it GW}+DMFT calculations
for the spectral properties of SrVO$_{3}$.
Fig.~\ref{GWDMFT}(a,b) displays the local projection of the spectral
function, while Fig.~\ref{fig:arpes} shows 
momentum dependent \t2g\ spectra in comparison with ARPES 
measurements\cite{PhysRevB.80.235104,PhysRevLett.109.056401}.
The global view on the spectral function in the full energy
range of valence and conduction band states, Fig.~\ref{GWDMFT}(a),
reveals an overall remarkable agreement with experiments. 
Indeed, GW+DMFT inherits from the GW calculation
the excellent agreement of the Sr-d states, both, in position
and shape, with BIS spectra,
and the improved agreement of the O-p ligand states with
photoemission.
The low-energy part of the spectrum is dominated by
the \t2g\ contribution, which, here, is profoundly modified with 
respect to pure {\it GW} results.
A renormalized
quasi-particle band disperses around the Fermi level~: 
At the $\Gamma$ point (see Fig.~\ref{fig:arpes})
the peak is located at about -0.5 eV --
this reveals (in agreement with ARPES)
a strong renormalization of the corresponding Kohn-Sham state 
which, at this momentum, has an energy of -1 eV.
At the X-point, the \t2g\ bands are no longer degenerate,
and surprisingly weakly renormalized xy/xz states are observed
at 0.9 eV, while the yz band is located at nearly the
same energy as at the $\Gamma$ point, again in agreement
with ARPES.  
At binding energies of -1.6 eV, ARPES witnesses a weakly dispersive
Hubbard band, whose intensity varies significantly as a function 
of momentum \cite{PhysRevB.80.235104}. 
In the {\it GW}+DMFT spectral function the Hubbard band -- absent in 
{\it GW} -- is correctly observed
at about -1.6 eV 
and its k-dependent intensity
variation (see Fig.~\ref{fig:arpes}) is indeed quite strong.
Previous LDA+DMFT calculations placed 
the lower Hubbard band at larger negative energies 
(see e.g.\ \cite{pavarini:176403}).
This is owing to the fact discussed above that when using a static Hubbard 
interaction, a value of 4--6 eV\cite{pavarini:176403,amadon:205112}, 
that is 
larger than the zero frequency limit of the 
{\it ab intio} $\mathcal{U}(\omega$$=$$0)$$=$$3.5$eV\cite{miyake:085122}, 
is needed to account
for the observed transfers of spectral weight.
As in DMFT with dynamic $\mathcal{U}$, 
{\it GW}+DMFT yields a good description of the Hubbard band and the spectral 
weight reduction at the same time, thanks to the
additional transfers of spectral weight due to the dynamical 
screening \cite{PhysRevB.85.035115,casula_effmodel,jmt_svo}.

At positive energies nonlocal self-energy effects are larger.
Interestingly, our k-integrated spectral function,
(see the dashed line in Fig.\ref{GWDMFT}(a) for the \t2g contribution
to the total (solid line) spectrum) does not display
a clearly separated Hubbard band. The reason is visible from the 
k-resolved spectra: the upper Hubbard band is located at around
2 eV, 
as expected from the location of the lower Hubbard band and
the fact that their separation is roughly given by the 
zero-frequency value of $U$. 
The peak around 2.7 eV that appears in the inverse photoemission
spectrum \cite{PhysRevB.52.13711} -- commonly interpreted as the
upper Hubbard band of \t2g character in the DMFT literature -- 
arises in fact from e$_{g}$ states located in this energy range.
The nonlocal
self-energy effects lead, in the unoccupied part of the
spectrum, to overlapping features from different k-points and
an overall smearing of the
total spectral function.

The Bose factor ansatz discussed above does not only provide us with an
efficient technique for solving the {\it GW}+DMFT equations.
It also 
allows for a transparent physical interpretation of the 
arising spectral properties.
Indeed the spectral representation of the bosonic 
renormalization factor $B(\tau)$ of Eq.~\ref{factorization-approximation}
(displayed in the lower panel of Fig.~\ref{fig:W}) 
is directly related
to the density of screening modes
$\Im\ \mathcal{U}(\omega)/\omega^2$.\cite{PhysRevB.85.035115}
In this way, we can trace back the {\it GW}+DMFT satellite 
at -4.5 eV
to the onset of p-\t2g\ excitations,
discussed above for $W$ and $U$. 
On the other hand, since the feature below 3~eV in $W$ is
absent in $U$ and $B$, the spurious {\it GW} peaks are consistently eliminated.
The strong peak at 15 eV is the 
well-known plasma excitation, seen
e.g.\ in electron energy loss spectra of SrTiO$_3$\cite{PhysRevB.62.7964}.

\subsubsection{Test of simplified schemes}

We now turn to the question of how to set up simplified schemes
that would still reproduce the results of the full \GW+DMFT 
calculations within the low-energy regime.
Besides the methodological interest, this study also allows
us to analyse more in detail the dominant effects leading
to corrections to the Kohn-Sham band structure.

As can be seen from the methodological section, the DMFT self-consistency
condition for the one-body quantities requires the local Green's function
to equal
\begin{eqnarray}
&&G^{loc} (i \omega) =\nonumber\\
 &&\sum_k [i \omega + \mu - H_0(k)
- \Sigma_{GW}^{non-loc}(k, i \omega) - \Sigma_{imp}(i \omega)]^{-1}
\nonumber
\\
\label{Eq:sc-ksum}
\end{eqnarray}
with $H_0=H_{LDA}-v^{xc}_{LDA}+\Sigma^x$ from Eq.~\ref{H0}, and   $\Sigma_{GW}^{non-loc}$ is the nonlocal part of the \GW\ t$_{2g}$ correlation self-energy $\Sigma_c=GW-GV$.
If 
the nonlocal correlation self-energy were purely static, that is $\omega$-independent, $\Sigma_{GW}^{non-loc}(k,\omega)=\Sigma_{GW}^{non-loc}(k)$, one could
construct an effective quasi-particle Hamiltonian that also comprises these correlation effects:

\begin{equation}%
H^{qp}(k)=H_0(k)+\Re\Sigma_{GW}^{non-loc}(k).%
\label{Hqp}%
\end{equation}
In Section \ref{GW} we have empirically shown (as seen before in Ref.~\onlinecite{jmt_pnict} for the iron pnictides) that nonlocal correlations are static
within the low-energy Fermi liquid regime.
This thus provides a justification for  using $H^{qp}$
for the construction of the free (of local correlation) propagator of the DMFT impurity.
$H^{qp}$ is the simplified one-shot analogue of the QS{\it GW} Hamiltonian $H^{\hbox{\tiny QS{\it GW}}}$ that was proposed in the context of the QS{\it GW}+DMFT formalism\cite{jmt_pnict}.
Then the DMFT self-consistency is much simpler since quantities are either frequency {\it or} momentum dependent, but not both, which drastically reduces
memory requirements. In QS{\it GW}+DMFT, an additional self-consistency on the {\it GW}-level is performed which circumvents the full  {\it GW}+DMFT self-consistency that
is computationally very demanding and has so far has only been achieved on the model level\cite{ayral_gwdmft, PhysRevB.87.125149}, and the simpler case of a
two-dimensional systems of adatoms on surfaces\cite{PhysRevLett.110.166401}.
Since, here, we only use a one-shot {\it GW} self-energy, without a self-consistency for its nonlocal contributions, we will refer to the scheme that uses Eq.~\ref{Hqp} as ``DMFT@nonlocal-{\it GW}''.

Here we will present a proof of principle that the described scheme 
yields excellent results for the properties that it was designed for. 
Of course, the simplified DMFT@nonlocal-{\it GW} scheme is not expected to give quantitatively accurate results {\it outside} the quasi-particle energy range. In particular the dispersion
of collective excitations will not be captured. However, their position in the local spectrum which is determined by the structure of the dynamic interaction $\mathcal{U}(\omega)$
is still meaningful as seen below.

\begin{figure*}%
\includegraphics[width=0.95\columnwidth,angle=0]{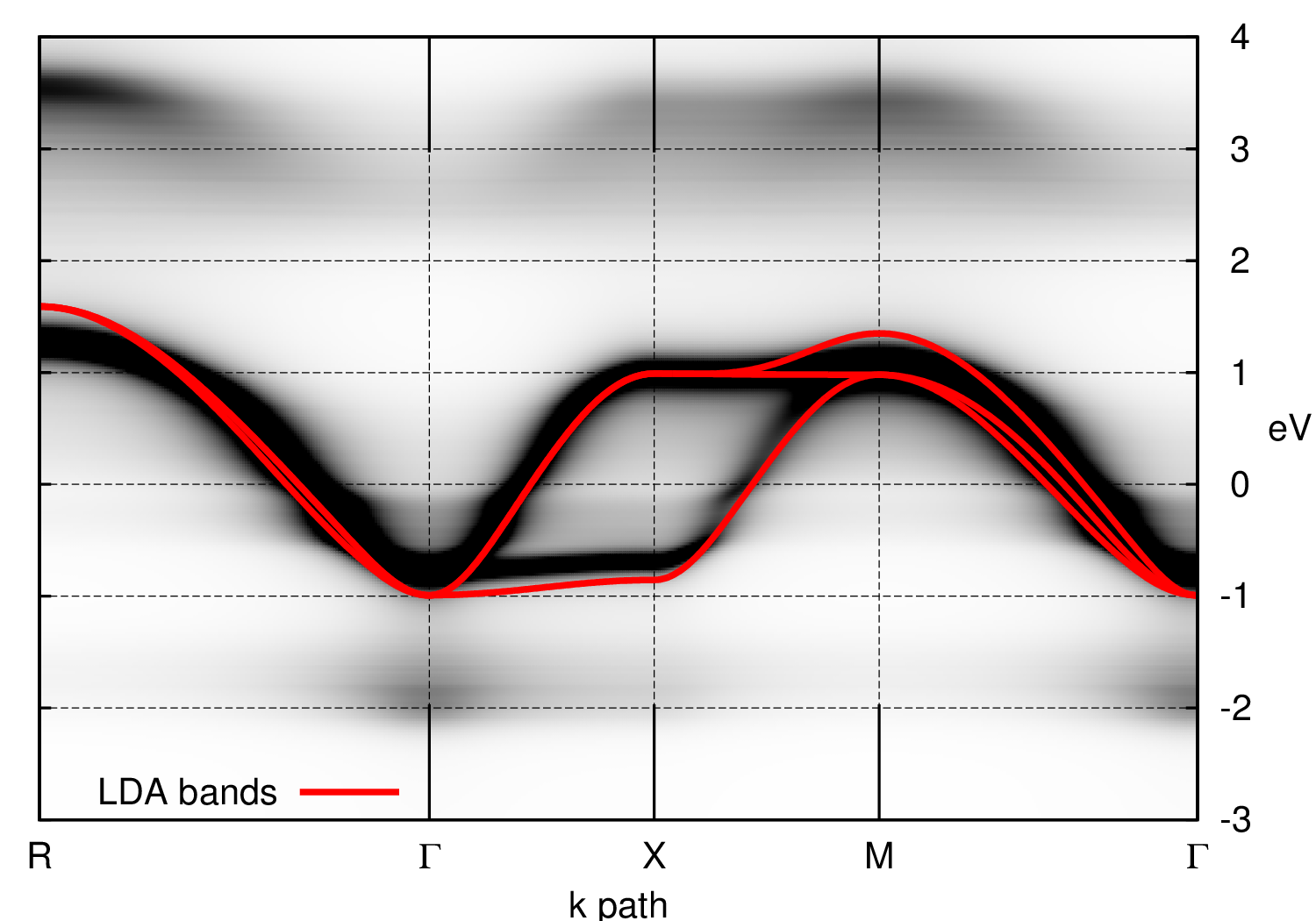} 
\includegraphics[width=\columnwidth,angle=0]{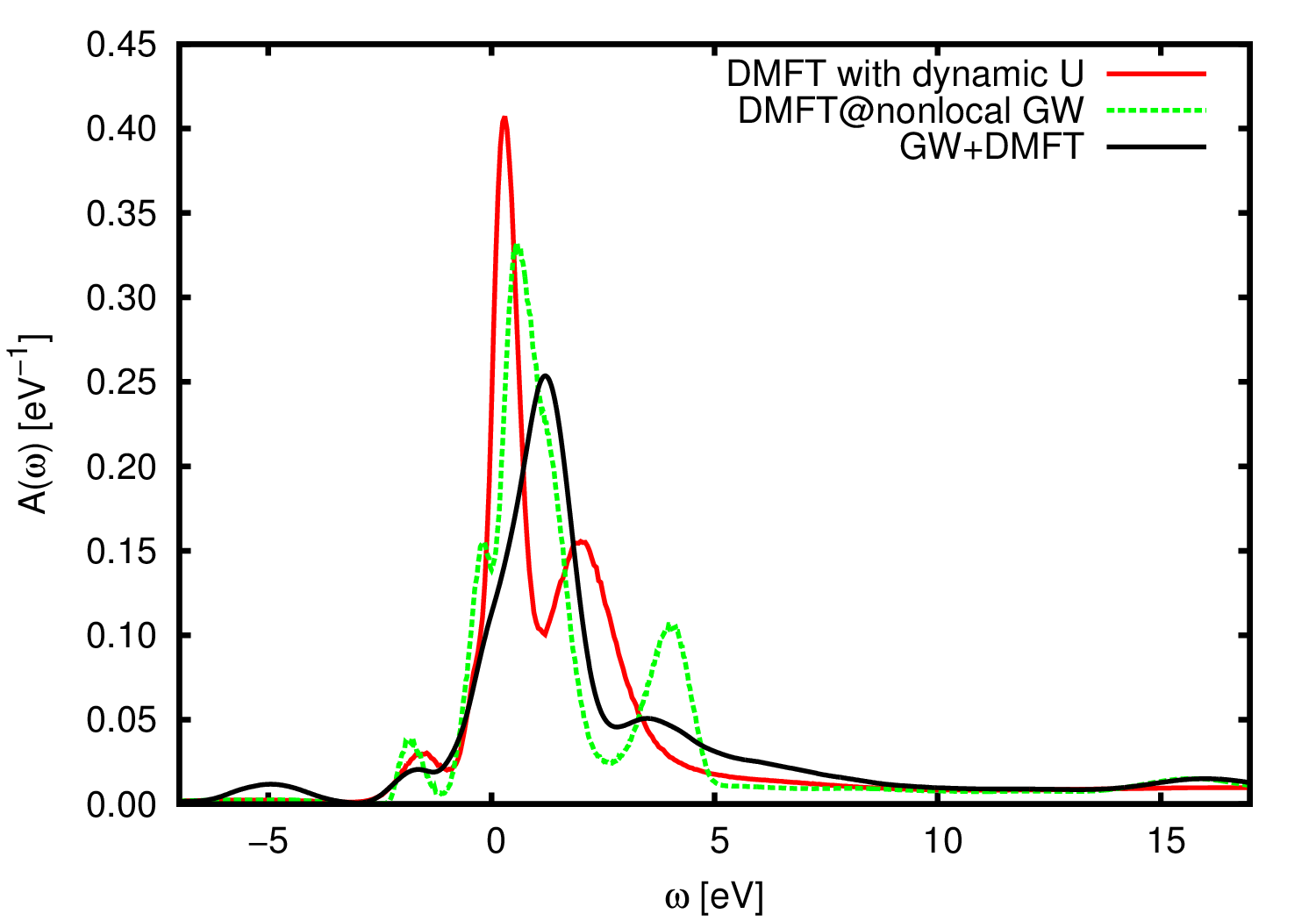}
\caption{(Color online) 
Spectra using the DMFT@nonlocal-{\it GW} approach: resolved along a momentum path (left) and
k-summed in comparison with the full {\it GW}+DMFT and LDA+$\mathcal{U}(\omega)$+DMFT.
} %
\label{Fig:nonloc+DMFT}%
\end{figure*}

For our current material, we can further simplify the approach. 
Indeed, for the fully degenerate t$_{2g}$ states, the local
self-energy is by symmetry a scalar (that is proportional
to the 3x3 unit matrix).
Eq.~(\ref{Eq:sc-ksum}) then reads for each of the three
orbital components
\begin{eqnarray}
&&G^{loc} (i \omega) =\nonumber\\
 &&\int d\epsilon D^{\hbox{\tiny eff}}(\epsilon) 
\frac{1}{i \omega + \mu - \epsilon
- \Sigma_{imp}(i \omega)}
\label{Eq:sc-ksum2}
\end{eqnarray}
where we have defined the density of states of the effective nonlocal-{\it GW} Hamiltonian $H^{qp}$ as
\begin{eqnarray}
D^{\hbox{\tiny eff}} (\epsilon) = 
-\frac{1}{\pi} Im \ Tr \sum_k [\epsilon - 
H^{qp}(k)
+ i 0^+]^{-1}\nonumber \\
\label{Eq:effdos}
\end{eqnarray}
This auxiliary quantity was discussed in the {\it GW} section and plotted in \fref{awgw}.
It contains all information on nonlocal correlations, and is double counting free when combined with
the DMFT self-energy.
Using this DOS with the cRPA $\mathcal{U}(\omega)$ in the DMFT methodology yields
the spectral function that is displayed in \fref{Fig:nonloc+DMFT}(a) along the usual k-path.
A comparison with \fref{akwgw} shows a remarkable agreement with the full {\it GW}+DMFT dispersion.
In panel (b) is further shown the local projection of this spectral function in comparison to the full {\it GW}+DMFT and the LDA+$\mathcal{U}(\omega)$+DMFT results.
Bearing in mind the different conditions for the analytical continuation necessary to obtain these spectra, the agreement between the DMFT@nonlocal-{\it GW}
and the full {\it GW}+DMFT is more than satisfactory: captured are the t$_{2g}$ bandwidth, the position of the lower Hubbard band, the satellite at +4eV,
and even the plasmon. 
The DMFT@nonlocal-{\it GW}, or the related
QS{\it GW}+DMFT\cite{jmt_pnict} approach, are thus promising approaches when a full {\it GW}+DMFT calculation
is too costly. 

\subsubsection{Further methodological remarks}

\begin{figure}%
\includegraphics[width=\columnwidth,angle=0]{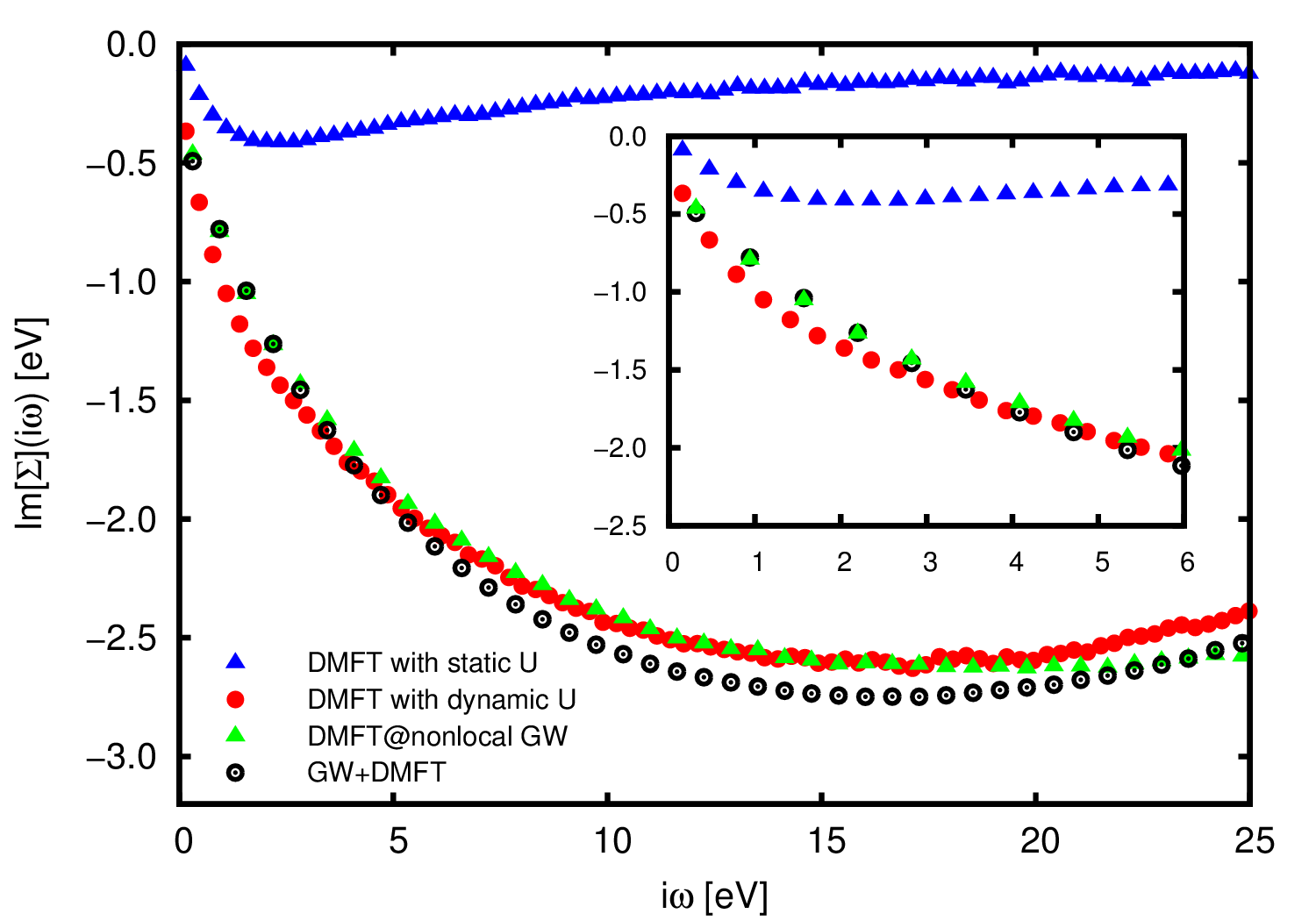}
\caption{(Color online) 
Comparison of local t$_{2g}$ self-energies on the Matsubara axis:
usual LDA+DMFT, LDA+DMFT with dynamical interactions, 
DMFT@nonlocal {\it GW} and local part of full {\it GW}+DMFT.
} %
\label{Fig:Sigmas}%
\end{figure}

Finally, we turn to a comparison of the contributions contained
in the different schemes, on the basis of the local part of the
Matsubara axis self-energies $\Sigma(i \omega)$, plotted in
Fig.~(\ref{Fig:Sigmas}).
The most striking feature in the comparison is the small amplitude
of the standard LDA+DMFT self-energy. This can be trivially understood
from the fact that only the partially screened value of the interaction,
the Hubbard $U$, enters into the description. This scheme does not
contain {\it any} information about the bare interaction -- in contrast
to all the other schemes, where it is recovered as the high-frequency limit.
This information does lead to much higher characteristic energy scales
in the schemes beyond LDA+DMFT, with self-energies living of the scale of the
plasma energy of $\sim$ 15 eV.

When comparing the shape of the self-energies at low-energies,
one can observe that the one for LDA+$\mathcal{U}(\omega)$+DMFT
is slightly steeper, leading to an overestimation of the 
mass renormalization. The self-consistency loop in the
full {\it GW}+DMFT scheme leads to a relaxation of the impurity
self-energy, and thus less important renormalization
effects.
The change is in fact quite substantial, leading to 
different quasi-particle weights corresponding to the different
self-energies: while the local $Z$ factor for the fully self-consistent 
{\it GW}+DMFT calculation is nearly 0.7, the LDA+DMFT calculation
with dynamical $U$ yields 0.5. A non-selfconsistent calculation
combining a local LDA+$\mathcal{U}(\omega)$+DMFT self-energy
with a nonlocal-{\it GW} self-energy could therefore be expected
to underestimate the bandwidth by a factor $0.7/0.5$. 
This may explain why a
recent paper for SrVO$_3$\cite{PhysRevB.88.235110} using such 
a non-selfconsistent 
``[LDA+$\mathcal{U}(\omega)$+DMFT]$_{local}$+[{\it GW}]$_{nonlocal}$''
approach finds a much narrower empty
band than our {\it GW}+DMFT calculations.
This puts strong constraints on the design of simplified
schemes: it
highlights the importance of having the nonlocal correlations 
present in the DMFT
self-consistency, as done in the DMFT@nonlocal-{\it GW} discussed above,
or also in the 
QS{\it GW}+DMFT scheme\cite{jmt_pnict}.

Finally, DMFT@nonlocal-{\it GW} and full {\it GW}+DMFT
are very close at low energies, as expected from the analysis
above, but start to deviate on a scale of a few eV where the
nonlocal self-energy correction on the real axis recovers some 
frequency-dependence. 
For obvious reasons, also a 
conceptually correct treatment of higher energy satellite 
features  
will require the use of the full {\it GW}+DMFT scheme.

\section{Summary}

We now come back to the list of physics questions
on our target compound, outlined in the introductory
section on SrVO$_3$:
\begin{itemize}
\item {\bf Reconciliation of results of DMFT
and cluster model calculations in the unoccupied
part of the spectra, and consistency with
{\it ab initio} $U$ values}
\\
Our finding of the IPES peak at 2.7 eV being
dominantly of e$_g$ character, rather than
being an upper Hubbard band of t$_{2g}$ character,
coincides with the interpretation of cluster
model calculations \cite{mossanek:033104}, thus
reconciling DMFT with the cluster model literature.
Cluster model calculations place an upper Hubbard
band of t$_{2g}$ character at about 2 eV, 
which would be consistent with the static value of
$\mathcal{U}(\omega=0)= 3.5$ eV.
The cluster calculations, however, do not have
access to the effects of the enhancement of the 
bare band dispersion by nonlocal exchange.
Our {\it GW}+DMFT calculations reveal that the latter
is in fact the dominant effect, preventing the
formation of a clearly separated sharp upper
Hubbard band.
\item {\bf The 2.7eV feature in BIS spectra}\\\
Related to the previous point, the photoemission
feature at 2.7 eV is {\it not} an upper Hubbard band,
but rather dominated by e$_g$ states. It would be
most interesting to perform orbital-resolved
inverse photoemission studies to confirm this
orbital assignment.
\item 
{\bf Position of O2p ligand states}
\\
The inclusion of the {\it GW} self-energy for the
``uncorrelated'' states, as explained in the
section on the orbital-separated {\it GW}+DMFT
scheme, introduces corrections on the O2p
ligand states, which are pushed down in energy,
improving the agreement with experiment.
We note, however, that the size of the correction
is not quite large enough, compared to experiment.
To some extent, this could have been expected:
indeed, we believe feedback effects of the
Coulomb interactions on the V-d states (and their
hybridization) to be important for determining
the O2p position. However, such effects would only
be included if an update of the {\it GW} part of the
calculation were also performed. This observation
thus opens important perspectives for further work.
\item {\bf Position of Sr-4d states}\\
The Sr-d states are pushed up by the {\it GW} self-energy.
The total O-p to Sr-d energetic distance is
enhanced by about 1.25eV. Comparison with the
experimental spectra shows that this correction
is excellent in the unoccupied part of the spectrum.
\end{itemize}
Most importantly, however, we identify a
substantial broadening of the unoccupied
bandwidth with respect to standard LDA+DMFT
calculations. Indeed, the nonlocal part
of the {\it GW} self-energy, when applied as
a correction to the LDA band structure,
leads to a widening by more than 40 percent.
When local correlations (within DMFT with
$\mathcal{U}(\omega)$) are
added, the corresponding renormalizations
re-narrow the unoccupied band roughly such
as to recover the original LDA bandwidth.
For this reason, while being similar to
the LDA+DMFT description for the occupied
states, our results suggest an entirely
new description for the unoccupied part
of the spectrum, calling for a reinvestigation
within techniques capturing empty states
properties (BIS, IPES, time-resolved ARPES or similar).

\section{Conclusions}

We have implemented the combined {\it GW}+DMFT scheme in a 
fully dynamical manner, by treating the {\it GW} part at
the one-shot level, but self-consisting over the DMFT
part.
Comparisons with pure LDA, pure {\it GW}, and LDA+DMFT calculations
with static and dynamic interactions allow to assess
the importance of the various features of these schemes, 
such as inclusion of dynamical screening, local and
nonlocal self-energy contributions, and self-consistency.

In particular our analysis suggests that at low-energies, 
the dynamical self-energy contributions of {\it GW} or combined
{\it GW}+DMFT schemes are strongly dominated by the local part,
and that the crucial nonlocal corrections are a purely
static correction to the LDA exchange correlation potential.
This is strongly encouraging in view of the accuracy of DMFT-based schemes
for correlated materials, and may allow for shortcuts when going beyond them (see e.g.\ the QS{\it GW}+DMFT scheme\cite{jmt_pnict}).

The calculated {\it GW}+DMFT spectral functions for SrVO$_3$
are in good agreement with available experimental
data for the occupied electronic states. In this
part of the spectra, the {\it GW}+DMFT scheme only leads
to a slight improvement over conventional LDA+DMFT
results (provided that in the latter the dynamics of 
the Hubbard $U$ is included). 

Very importantly, however, our {\it GW}+DMFT results also
suggest, that the {\it unoccupied}
band structure is not well described by many-body
calculations based on LDA-derived one-body Hamiltonians. 
Indeed, broadening by the Fock
exchange term is substantial; 
the appropriate bare band structure for a DMFT-based
electronic structure calculation should be wider
by about 40 $\%$ than the corresponding LDA bands,
so that the final dispersion {\it after the many-body
calculation} is eventually comparable again to the 
LDA one.

The mechanism leading to these corrections is quite
general: it is based on the simple observation that
the exchange-correlation potential of DFT provides
a much better approximation to (screened) exchange
for occupied electronic states than for empty ones.
This -- quite generally -- suggests
that -- despite their successes in describing occupied electronic
states -- {\it many-body techniques based on
LDA Hamiltonians are inappropriate for
describing unoccupied states} of correlated transition
metal oxides. This is
in particular true for the combined LDA+DMFT scheme.

These findings urgently call for experimental
studies of correlated oxides by techniques suitable for measuring
empty electronic states. Candidates could be
bremsstrahl-isochromatography (BIS)/inverse
photoemission, time-resolved ARPES, or more indirect
probes such as resonant inelastic x-ray scattering
(RIXS), optical spectroscopy, or x-ray absorption (XAS).

\section{Acknowledgements}
We acknowledge useful discussions with 
T. Ayral, A. Georges, M. Imada, A.I. Lichtenstein, 
C. Taranto, K. Held, as well as a collaboration
with F. Aryasetiawan at the early stage of Ref.~\onlinecite{jmt_svo}.
This work was supported by the French ANR under projects SURMOTT
and PNICTIDES and IDRIS/GENCI under projects 139313 and 096493. 
JMT acknowledges support from the
European Research Council under the European Union's Seventh 
Framework Programme (FP/2007-2013)/ERC
through grant n.\ 306447.

\subsection{Appendix: The {\it GW}+DMFT Equations}
\label{gwdmfteqn}

As discussed in Sec.~\ref{unified}, the {\it GW}+DMFT scheme as formulated in 
Refs.~\onlinecite{PhysRevLett.90.086402,gwdmft_proc1, gwdmft_proc2}
can be derived as a stationary point $(G, W)$
of the Almbladh free energy functional \cite{Almbladh_1999} 
after approximating the correlation part of
this functional by a combination of local and nonlocal terms
stemming from DMFT and {\it GW}, respectively.

For reference, we here review the equations derived from this construction,
leading to an
iterative loop which determines $\cG$ and $\cU$ self-consistently
(and, eventually, the full self-energy and polarization operators):
\begin{itemize}
\item The impurity problem (\ref{Simp}) is solved, for a given choice of $\cG_{LL'}$ and
$\cU_{\a\b}$: the ``impurity'' Green's function
\begin{equation}
G_{imp}^{LL'}\equiv - \langle T_\tau c_L(\tau)c^+_{L'}(\tau')\rangle_S
\label{Eq:Gimp}
\end{equation}
is calculated, together with the impurity self-energy
\begin{equation}
\Sigma^{xc}_{imp}\equiv\delta\Psi_{imp}/\delta G_{imp}=\cG^{-1}-G_{imp}^{-1}.
\end{equation}
The two-particle correlation function
\begin{equation}
\chi_{L_1L_2L_3L_4}=\langle :c^\dagger_{L_1}(\tau)c_{L_2}(\tau):
:c^\dagger_{L_3}(\tau')c_{L_4}(\tau'):\rangle_S 
\end{equation}
must also be evaluated.
\item
The impurity effective interaction is constructed as follows:
\begin{equation}\label{eff_inter}
W_{imp}^{\a\b} = \cU_{\a\b} -
\sum_{L_1\cdots L_4}\sum_{\gamma\delta} \cU_{\a\gamma}
O^{\gamma}_{L_1L_2}
\chi_{L_1L_2L_3L_4} [O^{\delta}_{L_3L_4}]^* \cU_{\delta\b}
\end{equation}
where 
$O_{L_1L_2}^\a\equiv\langle\phi_{L_1}\phi_{L_2}|B^\a\rangle$
is the overlap matrix between two-particle states and products
of one-particle basis functions.
The polarization operator of the impurity problem is then obtained as:
\begin{equation} 
P_{imp}\equiv -2\delta\Psi_{imp}/\delta W_{imp} = \cU^{-1}-W_{imp}^{-1},
\end{equation} 
where all  
matrix inversions are performed in the two-particle basis
$B^\a$
(see the discussion in \cite{gwdmft_proc1, gwdmft_proc2}).
\item
From Eqs.~(\ref{Sig_a}) and (\ref{P_a})
the full $\vk$-dependent Green's function $G(\vk,\iomn)$ and
effective interaction $W(\vq,\inun)$ can be constructed. The self-consistency condition
is obtained, as in the usual DMFT context, by requiring that the on-site
components of these quantities coincide with $G_{imp}$ and $W_{imp}$. In practice, this
is done by computing the on-site quantities
\begin{eqnarray}\label{Glocal}
G^{loc}(\iomn) &=& \sum_\vk [\GH^{-1}(\vk,\iomn) - \Sigma^{xc}(\vk,\iomn) ]^{-1}\\
\label{Wlocal}
W^{loc}(\inun) &=& \sum_\vq [V_{\vq}^{-1} - P(\vq,\inun)]^{-1}
\end{eqnarray}
and using them to update the
Weiss dynamical mean field ${\cal G}$
and the impurity model interaction ${\cal U}$ according to:
\begin{eqnarray}\label{update}
\cG^{-1} = {G^{loc}} ^{-1} + \Sigma^{xc}_{imp}
\\
\label{updateU}
\cU^{-1} = {W^{loc}} ^{-1} + P_{imp}
\end{eqnarray}
\end{itemize}
The set of equations (\ref{Eq:Gimp}) to (\ref{updateU}) 
(including (12) and (13))
is iterated until self-consistency.

This in fact means that, conceptually, there are two levels of
self-consistency: the one over local quantities, for 
a given GW calculation, and, eventually, also the update
of non-local quantities by recalculation the GW self-energies
and polarisation. In real materials calculations, this 
full self-consistency has been only performed once so far,
namely in the relatively simple case of a single-orbital
system \cite{PhysRevLett.110.166401}. Here, we restrict
ourselves to self-consistency at the DMFT level for a 
given GW calculation, as discussed in the methodological
sections above.


%

\end{document}